\documentclass[usenames,dvipsnames]{simplenote}    % Specifies the document style.
\newlength{\figurewidth}
\newlength{\subfigurewidth}
\newcommand{\of}[1]{\left(#1\right)}    % 'function of', i.e. (x) with proper parenthesis

\usetikzlibrary{arrows, shapes, calc, positioning}
\tikzstyle{base} = [text=white, text centered, minimum width=1em, minimum height=1em]
\tikzstyle{startstop} = [base, ellipse, fill=CernRed!90, draw=CernRed!200]
\tikzstyle{io} = [base, trapezium, trapezium left angle=70, trapezium right angle=110, fill=CernNiceBlue, draw=CernBlue!200]
\tikzstyle{process} = [base, rectangle, rounded corners, fill=AtlasOrange!90, draw=AtlasOrange!200]
\tikzstyle{test} = [base, signal, signal to=east and west, fill=GreenBellPepper!90, draw=GreenBellPepper!200]
\tikzstyle{arrow} = [->, >=latex]

\newcommand{\myover}[2]{\overset{\mathmakebox[\widthof{\ensuremath{#2}}]{#1}}{#2}}

\newcommand{\overeq}[1]{\myover{#1}{=}}

\newcommand{\ReOf}[1]{\Re \left[ #1 \right]}
\newcommand{\ImOf}[1]{\Im \left[ #1 \right]}
\newcommand{\nosign}[1][-]{\phantom{#1}}

\newcommand{\twocases}[5]{
\newsavebox{\mycases}% Store case "title" and brace
\begin{align}
  \sbox{\mycases}{$\displaystyle #1 \left\{\begin{array}{@{}c@{}}\vphantom{#2}\\\vphantom{#4}\end{array}\right.\kern-\nulldelimiterspace$}
  \raisebox{-.5\ht\mycases}[0pt][0pt]{\usebox{\mycases}}#2 #3\\
     #4 #5
\end{align}
}

\newcommand{\parlr}[1]{\left. \partial #1 \right.}
\newcommand{\pardelta}{\parlr{\delta}}
\newcommand{\parx}[1][]{\frac{\partial #1}{\partial x}}
\newcommand{\pary}[1][]{\frac{\partial #1}{\partial y}}

\makeatletter
\renewcommand*{\@textcolor}[3]{%
  \protect\leavevmode
  \begingroup
    \color#1{#2}#3%
  \endgroup
}
\makeatother
\newcommand{\gray}[1]{\textcolor{lightgray}{#1}}

\newcommand{\lrangle}[1]{\langle #1 \rangle}

\newcommand{\AC}[1]{\mathopen{}\mathclose\bgroup\left.#1^\text{AC}\aftergroup\egroup\right.}  % remove spacing from \left. and \right.

\usepackage[normalem]{ulem}
\makeatletter
\newcommand{\culine}[1]{\bgroup\markoverwith
{\textcolor{#1}{\rule[-0.3ex]{2pt}{0.4pt}}}\ULon}
\newcommand{\spacedcoloruline}[1]{\bgroup\markoverwith
{\textcolor{#1}{\rule[-0.8ex]{2pt}{0.4pt}}}\ULon}
\newcommand{\dcoloruline}[2]{\bgroup \UL@setULdepth
 \markoverwith{\lower\ULdepth\hbox
   {\kern-.03em\vbox{\color{#1}\hrule width.2em\kern1.2\p@\color{#2}\hrule}\kern-.03em}}%
 \ULon}
\makeatother
\newcommand\fracvphan{\vphantom{\frac{1}{2}}}
\newcommand{\colorjxy}[1]{\dcoloruline{cern}{AtlasRed}{\fracvphan#1}}
\newcommand{\colorjx}[1]{\spacedcoloruline{cern}{\fracvphan#1}}
\newcommand{\colorjxx}[1]{\dcoloruline{cern}{cern}{\fracvphan#1}}
\newcommand{\colorjy}[1]{\spacedcoloruline{AtlasRed}{\fracvphan#1}}
\newcommand{\colorjyy}[1]{\dcoloruline{AtlasRed}{AtlasRed}{\fracvphan#1}}
\documentversion{v1.0}
\date{20.01.2023}
\title{On the derivation of Amplitude Detuning and Chromaticity Formulas for Particle Accelerators}
\author{J. Dilly, M. le Garrec}


\begin{document}
\maketitle 

\begin{abstract}
        This note aims to provide a complete and comprehensible derivation of the contributions 
        to amplitude detuning in first order from multipoles up to dodecapolar order.
        In addition, the principle of feed-down is explored and example terms are given.
        The influence of an AC-Dipole is also included in the amplitude detuning. 
        The derivation starts from very basic principles and maths useful for the derivation,
        such as Taylor and Binomial expansion is given at the beginning of the note. 
\end{abstract}

\tableofcontents
\newpage

\section{Motivation}
Contributions from different magnetic field orders to the Hamiltonian of the accelerator system are the bread and butter 
of any accelerator physicist concerned with the accelerator optics.
While everyone should derive the form of these contribution at least once in their life by oneself - 
drawing Pascal triangles and calculating the binomial coefficients of $(x + iy)^n$ - a collection of all 
those formulas is hard to come by.
This note takes care of this circumstance and provides the reader not only with the final results of 
a multitude of parameters related to the Hamiltonian, but also tries to be thorough in the derivation of these relations,
so that any reader may easily follow the steps that lead us to the equations presented.

There are additional relations presented, which are generally known within the community, 
but whose specific formulas are hard to come by if not - to the best search capacity of the author - non-existent.
Examples for those are the specific relations of dodecapole fields to the second order Amplitude Detuning in \cref{sec:AmpDetNormalDodecapoles}
and contributions to chromaticity split by field order in \cref{sec:Chromaticity}.

Wholly new to the reader will be probably the change of Second Order Amplitude Detuning under the influence of driven excitation
with an AC-Dipole, as layed out in \cref{sec:AmpDetDecapolesACDipole}.

\notebox{
In this report the \textbf{indices for the magnetic fields components as well as for the field strengths
begin at 1, indicating a dipole field}. 
This is in contrast to MAD-X, where the dipole field strength is assigned  the index 0.
}

\notebox{
There are two different usages of $J$ in this note. 
One is a stand-in for the skew magnetic field component (in MAD-X called $KS$) and has an integer subscript showing the field order.
The other one represents the action, the invariant of linear motion. The action will have a plane - $x, y$ or $z$ - as subscript.
}

\section{Useful Maths}

This section contains useful relations which will be referenced to later.
\subsection{Real and Imaginary Parts}
With $x, y \in \mathbb{R}$ and $z \in \mathbb{C}$ we can note the relations

\begin{align}
    \ReOf{iz} = \ReOf{i(x+iy)} = \ReOf{ix-y} = -\ImOf{z} \label{eq:realOfiz} \\ 
    \ImOf{iz} = \ImOf{i(x+iy)} = \ImOf{ix-y} = \phantom{-}\ReOf{z} \label{eq:imagOfiz}  
\end{align}

\subsection{Binomial Expansion} %-----------------------------------------------
\label{sec:BinomialExpansion}
The binom $(x + y)^n$ can be expanded as

\begin{equation}
    \label{eq:binomiExpansion}
    (x + y)^n = \sum\limits_{k=0}^{n} \binom{n}{k} x^{k} \; y^{n-k} \;.
\end{equation}

\subsection{Taylor Expansions} %-------------------------------------------------
\subsubsection{Taylor Expansion in 2D}
From \cite{FeldmanMultivariableCalculus2021} Eq. (2.6.18) repeated here for reference:

\begin{align}
    \label{eq:2dTaylor}
    f(x + \Delta x, y + \Delta y)  \quad \overeq{Taylor} \quad
    \sum \limits_{\substack{l,m > 0 \\ l+m \leq n}}
    \frac{1}{l! \; m!} \frac{\partial^{l+m}}{\left. \partial x\right.^l \; \left. \partial y\right.^m} f(x, y) (\Delta x)^l (\Delta y)^m
\end{align}

\subsubsection{Taylor Expansion of \texorpdfstring{$z^n$}{z**n}}
\label{sec:TaylorExpansionZn}
With the $q^{th}$ derivative of the complex function $f(z) = z^n$ with $z \in \mathbb{C}$ 
\begin{equation}
    \label{eq:complexFunctionAndDerivative}
    f(z) = z^n \quad \overset{q^{\text{th}} \text{derivative}}{\Rightarrow} \quad \frac{\partial^q}{\parlr{z}^q} f(z) = \begin{cases}
        \frac{n!}{(n-q)!}z^{n-q} &  q \leq n\\ 
        0 & q > n
    \end{cases}
\end{equation}
the general Taylor expansion is
\begin{align}
    \label{eq:complexTaylor}
    f(z+\Delta z) \quad &\overeq{Taylor} \quad
                  \sum_{q=0} \frac{1}{q!} \frac{\partial^q}{\parlr{z}^q}f(z) \Delta z^q  \nonumber \\
                  &\overeq{\cref{eq:complexFunctionAndDerivative}} \quad
                  \sum_{q=0}^n \frac{n!}{q! \; (n-q)!} z^{n-q}\Delta z^q \;. 
\end{align}

\subsection{Expectation Values}
\label{sec:ExpectationValue}

The magnetic terms of the Hamiltonian (see~\cref{sec:HamiltonianTerms}) often depend on the phase $\phi_z$ at the contributing element.
As $\phi_z(s)$ describes not the phase advance but the actual phase at $s$, 
its value changes per turn (unless the system is in a resonant state).
Measurements of any phase dependent observable $f(\phi_x, \phi_y)$ over a reasonable amount of turns 
are therefore governed by the expectation value
\begin{equation}
    \label{eq:phaseExpectationDefinition}
    \langle f(\phi_x, \phi_y) \rangle = \frac{1}{4 \pi^2} \oint \oint f(\phi_x, \phi_y) \; d\phi_x d\phi_y \; .
\end{equation}

This consideration is in general not valid close to resonances, as here only certain parts of the phase-space might be covered.

\subsubsection{Expectation Values of Cosines}
Of use in this regard are the expectation values for $\langle cos(x)^{n} \rangle$, as given in 
Eq.(22) in \cite{WhiteDirectAmplitudeDetuning2013}. 
Shown here is their derivation.
In the following $k, n \in \mathbb{N}$.

We need to first define some relations:
\begin{align}
    \label{eq:binomiEix}
    (e^{ix} + e^{-ix})^{n} \quad &\overeq{\cref{eq:binomiExpansion}} \quad
    \sum\limits_{k=0}^{n} \binom{n}{k} \of{e^{ix}}^{k} \of{e^{-ix}}^{n-k} \nonumber\\
     &= \quad \sum\limits_{k=0}^{n} \binom{n}{k} e^{kix} \; e^{-inx} \; e^{ikx} \nonumber\\
     &= \quad \sum\limits_{k=0}^{n} \binom{n}{k} e^{i(2k-n)x} 
\end{align}
\begin{equation}
    \label{eq:intEix}
    \int_0^{2\pi}  e^{ikx} \; dx 
    =
    \left\{
    \begin{aligned}
    \int_0^{2\pi}  e^{ikx} \; dx &= \left. \frac{e^{ikx}}{ik} \right|_0^{2\pi} &= \frac{e^{ik2\pi} - 1}{ik} &= 0  &, k \neq 0\\
    \int_0^{2\pi}  e^{ikx} \; dx &=  \int_0^{2\pi}  1 \; dx &&= 2\pi &, k = 0 
    \end{aligned}
    \right\} = 2\pi\delta_{k}
    % && \Biggl| \; \substack{}
\end{equation}
\begin{align}
    \label{eq:intCos2n}
    \int_0^{2\pi} \cos^{2n}(x) \; dx &= \int_0^{2\pi} \of{ \frac{e^{ix} + e^{-ix}}{2} }^{2n} dx  
        &&\Biggl| \; \substack{
                    \text{\cref{eq:binomiEix}}, \\
                    \text{with } n \; \mapsto \; 2n
                    }
        \nonumber \\
        &=  \frac{1}{2^{2n}} \sum\limits_{k=0}^{2n} \binom{2n}{k} \int_0^{2\pi}  e^{i(2k-2n)x} \; dx
        &&\Biggl| \; \substack{
                    \text{\cref{eq:intEix}}, \\
                    \text{with } k \; \mapsto \; (2k - 2n) \\
                    \Rightarrow \; k \; \overset{!}{=} \; n 
                    }
        \nonumber \\
        &=  \frac{2\pi}{2^{2n}} \binom{2n}{n}
\end{align}
\begin{align}
    \label{eq:intCos2n1}
    \int_0^{2\pi} \cos^{2n+1}(x) \; dx &= \int_0^{2\pi} \of{ \frac{e^{ix} + e^{-ix}}{2} }^{2n+1} dx  
        &&\Biggl| \; \substack{
                    \text{\cref{eq:binomiEix}}, \\
                    \text{with } n \rightarrow 2n
                    }
        \nonumber \\
        &=  \frac{1}{2^{2n}} \sum\limits_{k=0}^{n} \binom{2n+1}{k} \int_0^{2\pi}  e^{i(2k-2n+1)x} \; dx
        &&\Biggl| \; \substack{
                    \text{\cref{eq:intEix}}, \\
                    \Rightarrow \; k \; \overset{?}{=} \; \frac{2n + 1}{2} \; \notin \; \mathbb{N} \\
                    \Rightarrow \; \text{all terms = 0}
                    }
        \nonumber \\
        &=  0
\end{align}

With these, we can now calculate the expectation value for cosines:
\begin{equation}
\label{eq:CosEvenExp}
    \langle \cos^{2n}(\phi_x) \rangle 
    \quad \overeq{\cref{eq:phaseExpectationDefinition}} \quad
    \frac{1}{4 \pi^2} \int_0^{2\pi} \int_0^{2\pi} \cos^{2n}(x) \; d\phi_x d\phi_y
    \quad \overeq{\cref{eq:intCos2n}} \quad
    \frac{1}{4 \pi^2} \int_0^{2\pi} \frac{2\pi}{2^{2n}} \binom{2n}{n} d\phi_y
    =
    2^{-2n} \of{ \begin{matrix} 2n \\ n \end{matrix} }
\end{equation}
\begin{equation}
\label{eq:CosOddExp}
    \langle \cos^{2n+1}(\phi_x) \rangle 
    \quad \overeq{\cref{eq:phaseExpectationDefinition}} \quad
    \frac{1}{4 \pi^2} \int_0^{2\pi} \int_0^{2\pi} \cos^{2n+1}(x) \; d\phi_x d\phi_y
    \quad \overeq{\cref{eq:intCos2n1}} \quad
    0
\end{equation}
More generally for the combination of cosine:
\begin{align}
\label{eq:MultiCosEvenExp}
\begin{split}
    \langle \cos^{2n}(\phi_x) \cos^{2m}(\phi_y) \rangle 
    \quad &\overeq{\cref{eq:phaseExpectationDefinition}} \quad
    \frac{1}{4 \pi^2} \int_0^{2\pi} \cos^{2n}(\phi_x) \; d\phi_x \int_0^{2\pi} \cos^{2m}(\phi_y) \; d\phi_y \\
    \quad &\overeq{\cref{eq:intCos2n}} \quad
    \frac{1}{4 \pi^2} \frac{2\pi}{2^{2n}} \binom{2n}{n} \frac{2\pi}{2^{2m}} \binom{2m}{m} \\
    &=
    2^{-2n} \; 2^{-2m} \of{ \begin{matrix} 2n \\ n \end{matrix} } \of{ \begin{matrix} 2m \\ m \end{matrix} }
\end{split}
\end{align}
and zero, if any of the exponents on the cosine is odd, due to \cref{eq:CosOddExp}.

Some example values that will be used later on:
\begin{equation}
    \label{eq:ExpCos}
    \langle \cos^2{\phi_z} \rangle = \frac{1}{2} , \qquad
    \langle \cos^4{\phi_z} \rangle = \frac{3}{8} , \qquad
    \langle \cos^6{\phi_z} \rangle = \frac{5}{16}
    \;.
\end{equation}

\section{Multipole Expansion}

The multipole expansion of a general magnetic field in x and y reads
\begin{equation}
    \label{eq:multipoleExp}
    B_y + iB_x = \sum^{\infty}_{n=1}\of{ B_n + iA_n } \of{ x + iy }^{n-1} \;.
\end{equation}
The normal and skew field gradients $B_n$ and $A_n$ can therefore be calculated from 
the complex field:
\begin{equation}
    \label{eq:multipoleGradients}
    B_n + iA_n = \frac{1}{(n-1)!} \left. \frac{\partial^{n-1}\left(B_y + iB_x\right)}{\partial (x + iy)^{n-1}}  \right\rvert_{x=0, y=0}
\end{equation}

The \textbf{main field component} (read: \textbf{only} component) of a \textbf{perfect} 2N-pole magnet
 is $B_N$ for a normal and $A_N$ for a skew magnet.
The multipole expansion of the magnetic field for a non-perfect magnet are based around this main component.
The coefficients $b_n$ and $a_n$ represent the normal and skew relative field errors at the reference radius $r_{ref}$:
\begin{equation}
    \label{eq:multipoleExpNormalAndSkew}
    B_y + iB_x = 
    \begin{cases}
        B_N \sum\limits^{\infty}_{n=1}\of{ b_n + ia_n }
       \of{ \frac{x + iy}{r_{ref}} }^{n-1} \; \text{, for normal magnets}\\
        A_N \sum\limits^{\infty}_{n=1}\of{ b_n + ia_n }
     \of{ \frac{x + iy}{r_{ref}} }^{n-1}  \; \text{, for skew magnets}
    \end{cases}   \;. 
\end{equation}
$b_n$ and $a_n$ are dimensionless but usually given in 'units' of $10^{-4}$.
Therefore the normal field component $B_n$ of 
an normal 2N-pole magnet is then
\begin{equation}
\label{eq:absoluteFromRelativeField}
    B_n\,[\si{\tesla\meter}^{1-n}] = B_N \cdot \frac{b_n}{r_{ref}^{n-1}} \;,
\end{equation}
and similar for the skew field component $A_n$ as well as the normal 
and skew field components for a skew 2N-pole magnet with the main 
field $A_N$.

\subsection{Field Normalization}
\begin{figure}[ht!]
    \centering
    \includegraphics[width=.5\textwidth]{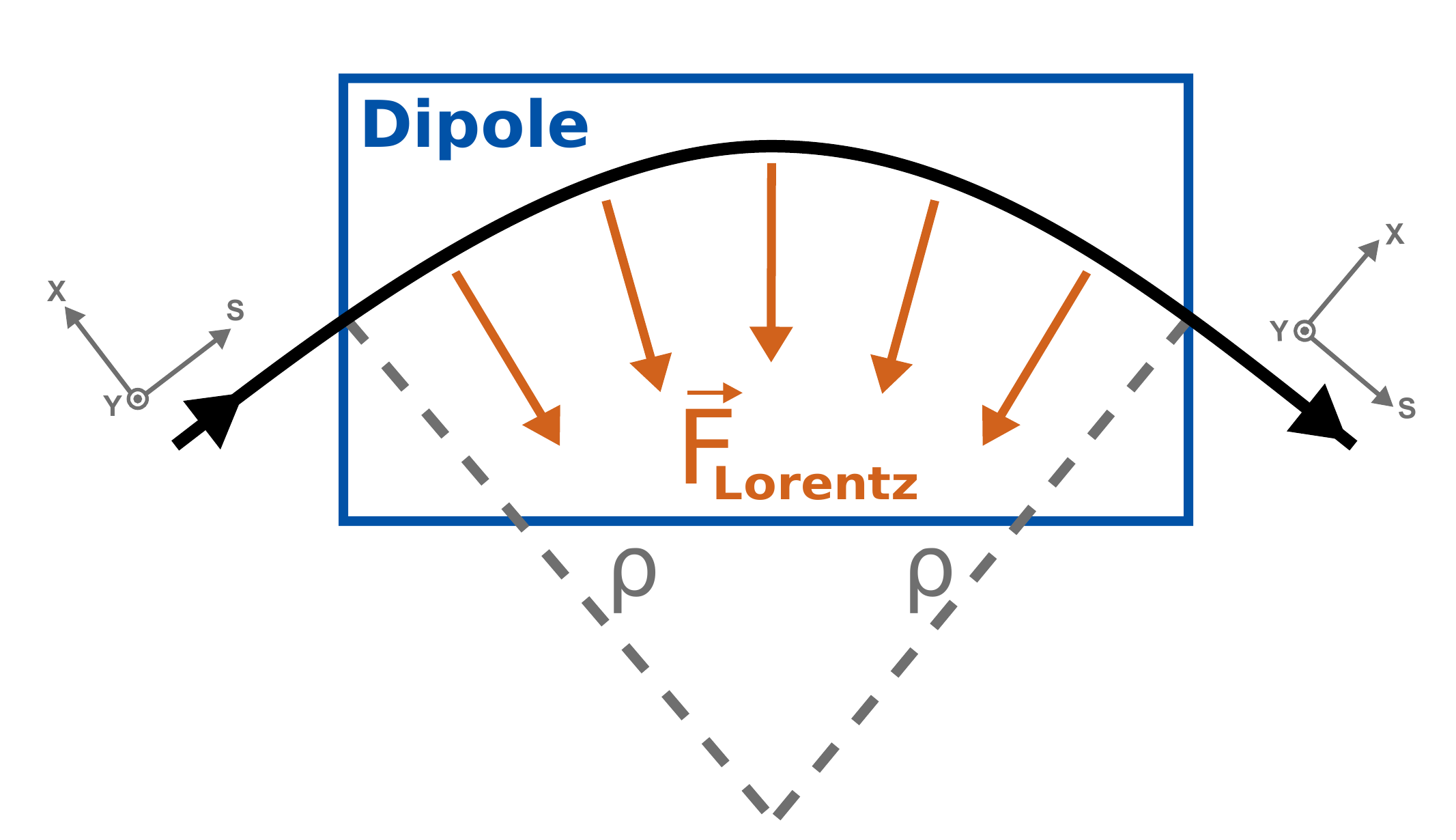}
    \caption{Schematic of the Lorentz force $\vec{F}_{Lorentz}$ and bending radius $\rho$ of a dipole with 
    magnetic field along $\vec{y}$ and a positive charged particle traversing along $\vec{s}$.}
    \label{fig:bendingRadiusDipole}
\end{figure}

The field gradients can furthermore be normalized to B$\rho$ (\textit{beam rigidity}),
the main dipole field and its bending radius (see \cref{fig:bendingRadiusDipole}):
\begin{equation}
    \label{eq:fieldCompNormalizationDerivative}
    K_n + iJ_n = \frac{1}{\mathrm{B}\rho} \left. \frac{\partial^{n-1}\left(B_y + iB_x\right)}{\partial (x + iy)^{n-1}}  \right\rvert_{x=0, y=0}
\end{equation}

In the following $q$ is the particle charge and $p$ is the beam momentum and
the Lorentz factor is defined as
\begin{equation}
    \label{eq:lorentzFactor}
    \gamma_0 = \frac{1}{\sqrt{1-\beta_0^2}} = \frac{1}{\sqrt{1-\frac{v^2}{c^2}}} \;.
\end{equation}
The beam rigidity can be calculated from the equality of the Lorentz force and the centripetal force:
\begin{align}
    & F_{\text{Lorentz}} &=& \qquad F_{\text{Centripetal}}  & \\
    \Rightarrow\; & qv \mathrm{B} &=& \qquad \frac{\gamma_0 m_0 v^2}{\rho} &= \frac{pv}{\rho} \\
    \Rightarrow\; & \mathrm{B} \rho &=& \qquad \frac{p}{q} &\;.
    \label{eq:BrhoToPQ}
\end{align}
With \cref{eq:fieldCompNormalizationDerivative,eq:BrhoToPQ}, the field strengths $K_n$, $J_n$ relate to $B_n$ and $A_n$:
\begin{equation}
    \label{eq:fieldCompNormalization}
    K_n \, [\si{\meter}^{-n}]= \frac{q}{p}  \, (n-1)! \, B_n,\quad J_n \, [\si{\meter}^{-n}]= \frac{q}{p} \, (n-1)! \, A_n  \;.
\end{equation}

\subsubsection{Fields of a Normal Dipole}
The main field component of a normal dipole $B_1$ according to 
\cref{eq:fieldCompNormalization} is:
\begin{equation}
    \label{eq:fieldCompDipole}
    B_1 = \frac{p}{q} \, K_1 \;. 
\end{equation}
Therefore, combining \cref{eq:fieldCompNormalization} and \cref{eq:absoluteFromRelativeField}
for a dipole reads:
\begin{align}
    B_n \quad&\overeq{\cref{eq:fieldCompNormalization}}\quad \frac{p}{q} \, \frac{1}{(n-1)!} \, K_n 
    \quad\overeq{\cref{eq:absoluteFromRelativeField}}\quad B_1 \cdot \frac{b_n}{r_{ref}^{n-1}} &&\Biggl| \; \substack{
         \cref{eq:fieldCompDipole}    \\
        } 
        \nonumber \\
    &\Leftrightarrow \frac{p}{q} \, \frac{1}{(n-1)!} \, K_n = \frac{p}{q} \, K_1 \cdot \frac{b_n}{r_{ref}^{n-1}} 
    &&\Biggl| \; \substack{
         \cdot \; \frac{q}{p}   \\
         \cdot \; (n-1)!
        } 
    \nonumber \\
    &\Leftrightarrow K_n = K_1 \cdot \frac{b_n}{r_{ref}^{n-1}} \cdot (n-1)!
\end{align}
which is a very useful relation to have.

\section{Hamiltonian Terms}
\label{sec:HamiltonianTerms}

The time dependent Hamiltonian for static magnetic field is
(\cite{WolskiBeamdynamicshigh2014} Eq. (2.22))
\begin{equation}
\label{eq:hamiltonianFullStaticTime}
    H = T - V = c \sqrt{\left|\bm p - q \bm A \right|^2 + m^2 c^2} + q\phi.
\end{equation}
After changing dependent variable to $s$ and in a curved trajectory reference system
(\cite{WolskiBeamdynamicshigh2014} Eq. (2.30)):
\begin{equation}
\label{eq:hamiltonianFullStatic}
    H = \frac{\delta}{\beta_0} - (1 + \frac{x}{\rho}) 
    \sqrt{\of{\delta + \frac{1}{\beta_0} }^2 - (p_x - a_x)^2 
    - (p_y - a_y)^2 - \frac{1}{\beta_0^2\gamma_0^2}} 
    - (1 + \frac{x}{\rho}) a_s \;,   
\end{equation}
with the energy deviation
\begin{equation}
    \label{eq:energyDeviation}
    \delta = \frac{E}{cp} - \frac{1}{\beta_0}     
\end{equation}
and the scaled vector potential
\begin{equation}
    \label{eq:scaledVectorPotential}
    \bm a = \frac{q}{p} \bm A \;.
\end{equation}

\subsection{Vector Potential}

With the \textbf{B} being the curl of \textbf{A} 
in the coordinates $(x,y,s)$ of the curved trajectory, see
\cite{WolskiBeamdynamicshigh2014} Eq. (3.36) - Eq. (3.38):
\begin{equation}
\textbf{B}(x, y, s) = \nabla \times \textbf{A}(x, y, s)  = 
    \begin{bmatrix} 
        \frac{\partial A_s}{\partial y} - 
        \frac{1}{1+ \frac{x}{\rho}} \frac{\partial A_y}{\partial s} \\
        \frac{1}{1+ \frac{x}{\rho}} \frac{\partial A_x}{\partial s} - 
        \frac{\partial A_s}{\partial x} -
        \frac{A_s}{\rho + x} \\
        \frac{\partial A_y}{\partial x} - \frac{\partial A_x}{\partial y} 
    \end{bmatrix}
    \overset{\overset{\text{gauge}}{A_x = A_y = 0}}{=}
    \begin{bmatrix} 
        \frac{\partial A_s}{\partial y}\\
        - \frac{\partial A_s}{\partial x} -
        \frac{A_s}{\rho + x} \\
        0
    \end{bmatrix} \;.
\end{equation}
For $n > 1$ (not a dipole) the path through the magnet center is straight and hence $\rho \rightarrow \infty$:
\begin{equation}
    \label{eq:fieldAsVectorNonDipole}
    B_y + iB_x = - \frac{\partial}{\partial x} A_s + i\frac{\partial}{\partial y} A_s \;.
\end{equation}
And hence
\begin{equation}
   \label{eq:asDefinition}
   A_s = - \ReOf{\sum_{n \geq 2} \frac{1}{n} (B_n + iA_n)(x+iy)^n} \;.
\end{equation}
As can be seen by calculating the derivative:
\begin{align}
    \label{eq:multipoleDExpansionFromAs}
    B_y + iB_x &= &-\frac{\partial}{\partial x} A_s 
                  &+ i\frac{\partial}{\partial y} A_s \nonumber\\
               &\overeq{\cref{eq:asDefinition}}
                &\qquad\ReOf{\sum_{n \geq 2} (B_n + iA_n)(x+iy)^{n-1}} 
                &- i \ReOf{\sum_{n \geq 2} i (B_n + iA_n)(x+iy)^{n-1}} \nonumber\\
               &\overeq{\cref{eq:realOfiz}} 
                &\qquad\ReOf{\sum_{n \geq 2} (B_n + iA_n)(x+iy)^{n-1}} 
                &+ i \ImOf{\sum_{n \geq 2} (B_n + iA_n)(x+iy)^{n-1}} \nonumber\\
               &= &\sum_{n \geq 2} (B_n + iA_n)(x+iy)^{n-1} & \;,
\end{align}
which is the multipole expansion of \cref{eq:multipoleExp} without the contributions from dipoles.
The magnetic field contribution to the Hamiltonian is then, 
according to \cref{eq:hamiltonianFullStatic,eq:scaledVectorPotential,eq:asDefinition}:
\begin{align}
   \label{eq:hamiltonianFieldPart}
   \begin{split}
   H &= \qquad - \frac{q}{p} \ReOf{\sum_{n \geq 2} \frac{1}{n} (B_n + iA_n)(x+iy)^n} \\
     &\overeq{\cref{eq:fieldCompNormalization}} \qquad
     -\ReOf{\sum_{n \geq 2} (K_n + iJ_n)\frac{(x+iy)^n}{n!}} 
   \end{split}
\end{align}

\subsection{Hamiltonian terms examples from different sources}
For different rotations and orders of magnetic field components, we can 
get the corresponding Hamiltonian, i.e. their contribution to the total Hamiltonian as:

\vspace{\baselineskip}
Normal Hamiltonian term:
\begin{equation}
    \label{eq:normalHamilStrength}
    N_n = \frac{1}{n!} \, K_n \, \ReOf{(x+iy)^{n}}
\end{equation}

Skew Hamiltonian term:
\begin{equation}
    \label{eq:skewHamilStrength}
    S_n = - \frac{1}{n!} \, J_n \, \ImOf{(x+iy)^{n}}
\end{equation}

As a reference for the reader, these terms are written out up to $6^{th}$ order in the next sections.

\subsubsection{First Order}
\begin{equation}
    \label{eq:n1O}
    N_1\of{x,y} = K_1 \, x
\end{equation}
\begin{equation}
    \label{eq:s1O}
    S_1\of{x,y} = - J_1 \, y
\end{equation}

\subsubsection{Second Order}
\begin{equation}
    \label{eq:n2O}
    N_2\of{x,y} = \frac{1}{2} \, K_2 \, \of{x^2 - y^2}
\end{equation}
\begin{equation}
    \label{eq:s2O}
    S_2\of{x,y} =  - J_2 \, xy
\end{equation}

\subsubsection{Third Order}
\begin{equation}
    \label{eq:n3O}
    N_3\of{x,y} = \frac{1}{3!} \, K_3 \, \of{x^3 - 3xy^2}
\end{equation}
\begin{equation}
    \label{eq:s3O}
    S_3\of{x,y} =  - \frac{1}{3!} \, J_3 \, \of{3x^2y - y^3}
\end{equation}

\subsubsection{Fourth Order}
\begin{equation}
    \label{eq:n4O}
    N_4\of{x,y} = \frac{1}{4!} \, K_4 \, \of{x^4 - 6x^2y^2 + y^4}
\end{equation}
\begin{equation}
    \label{eq:s4O}
    S_4\of{x,y} =  - \frac{1}{4!} \, J_4 \, \of{4x^3y - 4xy^3}
\end{equation}

% Fifth Order --------------------------------------------------------------------------------------

\subsubsection{Fifth Order}
\begin{equation}
    \label{eq:n5O}
    N_5\of{x,y} = \frac{1}{5!} \, K_5 \, \of{x^5 - 10x^3y^2 + 5xy^4}
\end{equation}
\begin{equation}
    \label{eq:s5O}
    S_5\of{x,y} =  - \frac{1}{5!} \, J_5 \, \of{5x^4y - 10x^2y^3 + y^5}
\end{equation}

% Sixth Order --------------------------------------------------------------------------------------

\subsubsection{Sixth Order}
\begin{equation}
    \label{eq:n6O}
    N_6\of{x,y} = \frac{1}{6!} \, K_6 \, \of{x^6 - 15x^4y^2 + 15x^2y^4 - y^6}
\end{equation}
\begin{equation}
    \label{eq:s6O}
    S_6\of{x,y} =  - \frac{1}{6!} \, J_6 \, \of{6x^5y - 20x^3y^3 + 6xy^5}
\end{equation}

% Feed-Down -------------------------------------------------------------------------------------------
\subsection{Feed-Down}
\label{sec:FeedDown}

A charged particle that traverses a magnetic field off-center with the horizontal and vertical offsets $\Delta x$ and $\Delta y$
experiences so called feed-down effects.
The effective magnetic field this particle sees can be described as a superposition of the main magnetic
field of the magnet, plus fields of lower order which scale with the particles offset.
Feed-down can be calculated via Taylor Expansion $x \mapsto x + \Delta x$ and $y \mapsto y + \Delta y$ 
from the Hamiltonian terms \cref{eq:hamiltonianFieldPart}. 

As a convention, terms that have the general coefficients as Hamiltonian terms of a certain order (as in 
\cref{eq:normalHamilStrength,eq:skewHamilStrength}), but contain $K_n$ and $J_n$ values of different order
are denoted by adding these in parenthesis in the superscript to the Hamiltonian terms name, 
e.g. $N_n^{(J_m)}$ is the normal Hamiltonian of order $n$ governed by the skew magnetic field strength $J_m$ of order $m$.

\subsubsection{General Terms up to Fourth Order Feed-Down}
From \cref{eq:normalHamilStrength,eq:skewHamilStrength} we can calculate the derivatives with respect to $x, y$
\begin{align}
    \parx{N_n^{(L_m)}} \quad 
    &\overeq{\cref{eq:normalHamilStrength}} \quad 
    \nosign \frac{1}{n!} \, L_m \, \ReOf{n(x+iy)^{n-1}}
    &= \nosign \frac{1}{(n-1)!} \, L_m \, \ReOf{(x+iy)^{n-1}}
    &= \nosign N_{n-1}^{(L_m)} \label{eq:dNndx}\\
    \pary{N_n^{(L_m)}} \quad 
    &\overeq{\cref{eq:normalHamilStrength}} \quad 
    \nosign \frac{1}{n!} \, L_m \, \ReOf{in(x+iy)^{n-1}}
    \quad &\overeq{\cref{eq:realOfiz}}
    -\frac{1}{(n-1)!} \, L_m \, \ImOf{(x+iy)^{n-1}}
    &= \nosign S_{n-1}^{(L_m)} \label{eq:dNndy}\\
    \parx{S_n^{(L_m)}} \quad 
    &\overeq{\cref{eq:skewHamilStrength}} \quad 
    -\frac{1}{n!} \, L_m \, \ImOf{n(x+iy)^{n-1}}
    &= -\frac{1}{(n-1)!} \, L_m \, \ImOf{(x+iy)^{n-1}}
    &= \nosign S_{n-1}^{(L_m)} \label{eq:dSndx}\\
    \pary{S_n^{(L_m)}} \quad 
    &\overeq{\cref{eq:skewHamilStrength}} \quad 
    -\frac{1}{n!} \, L_m \, \ImOf{in(x+iy)^{n-1}}
    \quad &\overeq{\cref{eq:imagOfiz}}
    -\frac{1}{(n-1)!} \, L_m \, \ReOf{(x+iy)^{n-1}}
    &= -N_{n-1}^{(L_m)} \label{eq:dSndy}
\end{align}
with $L_m$ as a stand-in for either $J_m$ or $K_m$.
The Taylor expansion (\cref{eq:2dTaylor}) of $N_n$ and $S_n$ therefore reads
\begin{equation}
    \label{eq:feedn6O}
    \begin{split}
        &N_n\of{x+\Delta x,y+\Delta y} = \; \\
                 &               N_n\of{x, y} \\
            + \; &               N_{n-1}^{(K_n)}\of{x, y}\Delta x + S_{n-1}^{(K_n)}\of{x, y}\Delta y \\
            + \; &\frac{1}{2} \, N_{n-2}^{(K_n)}\of{x, y}\of{\Delta x^2 - \Delta y^2} \;
            + \;                 S_{n-2}^{(K_n)}\of{x, y}\Delta x \Delta y \\
            + \; &\frac{1}{6} \, N_{n-3}^{(K_n)}\of{x, y}\of{\Delta x^3 - 3 \, \Delta x \Delta y^2} \; 
            + \;  \frac{1}{6} \, S_{n-3}^{(K_n)}\of{x, y}\of{3 \, \Delta x^2 \Delta y - \Delta y^3}  \\ 
            + \; &\frac{1}{24}\, N_{n-4}^{(K_n)}\of{x, y}\of{\Delta x^4 - 6 \, \Delta x^2 \Delta y^2 + \Delta y^4} \;
            + \;  \frac{1}{6} \, S_{n-4}^{(K_n)}\of{x, y}\of{\Delta x^3 \Delta y - \Delta x \Delta y^3}  \\ 
            + \; &\frac{1}{120} \, N_{n-5}^{(K_n)}\of{x, y}\of{\Delta x^5 - 10 \Delta x^3 \Delta y^2 + 5 \Delta x \Delta y^4} 
            +     \frac{1}{120} \, S_{n-5}^{(K_n)}\of{x, y}\of{5 \Delta x^4 \Delta y - 10 \Delta x^2 \Delta y^3 + \Delta y^5}  \\ 
            + \; &h.o.t.
    \end{split}
\end{equation}
\begin{equation}
    \label{eq:feeds6O}
    \begin{split}
        &S_n\of{x+\Delta x,y+\Delta y} =  \\
                 &               S_n\of{x, y}\\
            + \; &               S_{n-1}^{(J_n)}\of{x, y}\Delta x - N_{n-1}^{(J_n)}\of{x, y}\Delta y\\
            + \; &\frac{1}{2} \, S_{n-2}^{(J_n)}\of{x, y}\of{\Delta x^2 - \Delta y^2} \; 
            - \;                 N_{n-2}^{(J_n)}\of{x, y}\Delta x \Delta y\\
            + \; &\frac{1}{6} \, S_{n-3}^{(J_n)}\of{x, y}\of{\Delta x^3 - 3 \,\Delta x \Delta y^2} \; 
            - \;  \frac{1}{6} \, N_{n-3}^{(J_n)}\of{x, y}\of{3 \, \Delta x^2 \Delta y - \Delta y^3} \\ 
            + \; &\frac{1}{24} \, S_{n-4}^{(J_n)}\of{x, y}\of{\Delta x^4 - 6 \, \Delta x^2 \Delta y^2 + \Delta y^4} \; 
            - \;  \frac{1}{6} \, N_{n-4}^{(J_n)}\of{x, y}\of{\Delta x^3 \Delta y - \Delta x \Delta y^3} \\ 
            + \; &\frac{1}{120} \, S_{n-5}^{(J_n)}\of{x, y}\of{\Delta x^5 - 10 \Delta x^3 \Delta y^2 + 5 \Delta x \Delta y^4} 
            -     \frac{1}{120} \, N_{n-5}^{(J_n)}\of{x, y}\of{5 \Delta x^4 \Delta y - 10 \Delta x^2 \Delta y^3 + \Delta y^5}  \\ 
            + \; &h.o.t.
    \end{split}
\end{equation}

\section{Amplitude Detuning}
\label{sec:AmplitudeDetuning}

The tune, expanded around beam center, reads in terms of emittance $\epsilon_{z} = 2J_z$ (see also~\cite{MacleanAmplitudeDetuningMeasurements2015}):
\begin{equation}
    \label{eq:tuneExpansion}
    Q_z(\epsilon_x, \epsilon_y) = Q_{z0} + \of{\frac{\partial Q_z}{\partial \epsilon_x}\epsilon_x + \frac{\partial Q_z}{\partial \epsilon_y}\epsilon_y } 
    + \frac{1}{2} \left(\frac{\partial^2 Q_z}{\partial \epsilon_x^2}\epsilon_x^2 
                  + 2\frac{\partial^2 Q_z}{\partial \epsilon_x \epsilon_y}\epsilon_x\epsilon_y 
                  + \frac{\partial^2 Q_z}{\partial \epsilon_y^2}\epsilon_y^2
    \right) + ... \; .
\end{equation}
Further, $\phi_z$ and $J_z$ are canonical coordinates in the Courant-Snyder coordinate system, and hence the Hamiltonian equations read:

\begin{equation}
    \label{eq:hamiltonianEqActionAngle}
    \frac{\partial \phi_z}{\partial s} = \frac{\partial H}{\partial J_z} 
    , \qquad 
    \frac{\partial J_z}{\partial s} = - \frac{\partial H}{\partial \phi_z} 
\end{equation}

With the definition of the tune as 
\begin{equation}
    \label{eq:tuneDefinition}
    Q_z = \frac{1}{2\pi} \oint \frac{\partial \phi_z}{\partial s} ds
\end{equation}
the change of the tune within one magnet of length $L$ therefore becomes
\begin{equation}
    \label{eq:tuneChange}
    \Delta Q_z = \frac{1}{2\pi} \int_L \frac{\partial H }{\partial J_z} ds \;,
\end{equation}
where $H$ is the contribution to the Hamiltonian of that magnet.
As we are interested in the effect over many turns, we actually need to calculate the 
expectation value (see~\cref{sec:ExpectationValue}):
\begin{equation}
    \label{eq:tuneChangeExpectation}
    \Delta Q_z = \frac{1}{2\pi} \int_L \frac{\partial \langle H \rangle }{\partial J_z} ds \;,
\end{equation}

\subsection{From Normal Octupoles}
\label{sec:AmpDetNormalOctupoles}

Using the Hamiltonian for the normal octupole (\cref{eq:n4O}) with
action-angle coordinates
\begin{equation}
    \label{eq:ActionAngle}
    z = \sqrt{2J_z\beta_z} \cos{\phi_z} \; ,
\end{equation}
we get
\begin{equation}
    \label{eq:n40CS}
    N_4 = \frac{1}{4!} K_4 \of{ 4J_x^2\beta_x^2 \cos^4{\phi_x} + 24J_x J_y\beta_x\beta_y \cos^2{\phi_x}\cos^2{\phi_y} + 4J_y^2\beta_y^2 \cos^4{\phi_y}} \; .
\end{equation}

When calculating the expectation value $\langle N_4 \rangle$ (see~\cref{eq:phaseExpectationDefinition}) from \cref{eq:n40CS}, 
we can apply $\langle \cos^2 \rangle$ from~\cref{eq:ExpCos}, which then yields
\begin{equation}
    \label{eq:n40ExpCalc}
    \langle N_4 \rangle = \frac{1}{16} K_4 \of{J_x^2\beta_x^2 + 4 J_x J_y\beta_x\beta_y + J_y^2\beta_y^2} \;.
\end{equation}

Further simplifications can be made under the assumption that $\phi_z$, $J_z$ and $\beta_z$ are approximately constant within one magnet. 
In the case of a normal octupole \cref{eq:tuneChange} then becomes
\begin{equation}
    \label{eq:tuneChangeN4}
    \Delta Q_{x,y}^{N_4} = \frac{1}{2\pi} \int_L \frac{\partial \langle N_4 \rangle }{\partial J_{x,y}} ds 
    \quad \overeq{\cref{eq:n40ExpCalc}} \quad
    \frac{1}{32\pi} K_4L 
    \frac{\partial}{\partial J_{x,y}}
     \of{J_x^2\beta_x^2 + 4 J_x J_y\beta_x\beta_y + J_y^2\beta_y^2}
\end{equation}

\begin{equation}
    \label{eq:deltaQx}
    \Delta Q_x^{N_4} = \frac{1}{16\pi} K_4L \of{\beta^2_xJ_x - 2 \beta_x \beta_y J_y}
\end{equation}
\begin{equation}
    \label{eq:deltaQy}
    \Delta Q_y^{N_4} = \frac{1}{16\pi} K_4L \of{\beta^2_yJ_y - 2 \beta_x \beta_y J_x}
\end{equation}

The first order detuning with the action $2J_z$ in \cref{eq:tuneExpansion} is therefore:
\begin{equation}
    \label{eq:ampdetDirectX}
    \frac{\partial Q_x^{N_4}}{\partial (2J_x)} =  \frac{K_4L}{32\pi} \, \beta^2_x
\end{equation}
\begin{equation}
    \label{eq:ampdetCross}
    \frac{\partial Q_x^{N_4}}{\partial (2J_y)} 
    \overset{\cref{eq:tuneChangeN4}}{=} 
    \frac{\partial Q_y^{N_4}}{\partial (2J_x)} = - \frac{K_4L}{16\pi} \, \beta_x \beta_y
\end{equation}
\begin{equation}
    \label{eq:ampdetDirectY}
    \frac{\partial Q_y^{N_4}}{\partial (2J_y)} = \frac{K_4L}{32\pi} \, \beta^2_y
\end{equation}

\subsubsection{With AC-Dipole} %--------------------------------------------------------------------
\label{sec:AmpDetOctupolesACDipole}
Introducing a driven oscillation into the system, for example via an AC-Dipole, the final transversal position $z$
of the free and the driven oszillation, as in \cite{WhiteDirectAmplitudeDetuning2013}:
\begin{equation}
    z \mapsto  z + \AC{z} \;.
\end{equation}
The forced oscillation can also be written in action-angle coordinates
\begin{equation}
    \label{eq:ActionAngleAC}
    \AC{z} = \sqrt{2\AC{J_z}\AC{\beta_z}} \cos{\AC{\phi_z}} \;,
\end{equation}
and so
\begin{equation}
    \label{eq:ActionAngleFreeAC}
    z = \sqrt{2J_z\beta_z} \cos{\phi_z} + \sqrt{2\AC{J_z}\AC{\beta_z}} \cos{\AC{\phi_z}} \;.
\end{equation}

As before, we need the expectation value of the hamiltonian from normal octupoles.
We need now the expectation values for $\phi_z$ and $\AC{\phi_z}$, so instead of the definition of \cref{eq:phaseExpectationDefinition},
we use 
\begin{equation}
    \label{eq:phaseExpectationDefinitionAC}
    \langle f(\phi_x, \phi_y, \AC{\phi_x}, \AC{\phi_y}) \rangle = \frac{1}{16 \pi^4} \iiiint \limits_0^{2\pi} f(\phi_x, \phi_y, \AC{\phi_x}, \AC{\phi_y}) \; d\phi_x d\phi_y d\AC{\phi_x} d\AC{\phi_y}\; ,
\end{equation}
so that the individual integrals are still normalized.
Terms with expectation values of zero according to~\cref{eq:CosOddExp} are greyed out in the second step.
In the final step, terms depending on $J_x$ are underlined blue, terms depending on $J_y$ red, and terms depending on neither are in grey.
\newcommand{\tmplineintro}{\phantom{=\quad \frac{1}{4!} K_4 \Bigg(}}
\begin{align}
\label{eq:n40ExpAC}
\begin{split}
    \langle \AC{N_4} \rangle 
    \quad &\overeq{\cref{eq:n4O}}\quad
    \langle \frac{1}{4!} K_4 \of{x^4 - 6 x^2 y^2 + y^4} \rangle \\
    \quad &\overeq{z \mapsto z + \AC{z}} \quad
    \frac{1}{4!} K_4 \Bigg(
                   \lrangle{x^4} \gray{ + 4 \lrangle{x^3} \lrangle{\AC{x}}} + 6 \lrangle{x^2}\lrangle{\AC{x}^2} \gray{+ 4 \lrangle{x}\lrangle{\AC{x}^3}} + \lrangle{\AC{x}^4} \\
    &\tmplineintro - 6 \left( \lrangle{x^2} \gray{+ 2 \lrangle{x} \lrangle{\AC{x}}} + \lrangle{\AC{x}^2} \right) \left( \lrangle{y^2} \gray{+ 2 \lrangle{y} \lrangle{\AC{y}}} + \lrangle{\AC{y}^2} \right)\\
    &\tmplineintro + \lrangle{y^4} \gray{ + 4 \lrangle{y^3} \lrangle{\AC{y}}} + 6 \lrangle{y^2}\lrangle{\AC{y}^2} \gray{+ 4 \lrangle{y}\lrangle{\AC{y}^3}} + \lrangle{\AC{y}^4}
    \Bigg) \\
    &\overeq{\cref{eq:CosOddExp}} \quad
    \frac{1}{4!} K_4 \Bigg( \lrangle{x^4} + 6 \lrangle{x^2}\lrangle{\AC{x}^2} + 4 \lrangle{x}\lrangle{\AC{x}^3} + \lrangle{\AC{x}^4} \\
    &\tmplineintro - 6 \left( \lrangle{x^2}\lrangle{y^2} + \lrangle{x^2}\lrangle{\AC{y}^2} + \lrangle{\AC{x}^2} \lrangle{y^2} + \lrangle{\AC{x}^2}\lrangle{\AC{y}^2} \right) \\
    &\tmplineintro \lrangle{y^4}  + 6 \lrangle{y^2}\lrangle{\AC{y}^2} + 4 \lrangle{y}\lrangle{\AC{y}^3} + \lrangle{\AC{y}^4}
    \Bigg) \\
    % &\myoverunder{\cref{eq:ExpCos}}{=}{\cref{eq:ActionAngleFreeAC}} \quad
    &\overeq{\substack{\cref{eq:ExpCos} \\ \cref{eq:ActionAngleFreeAC}}} \quad
    \frac{1}{4!} K_4 \Bigg( 
        \colorjx{\frac{3}{2} J_x^2 \beta_x^2 + 6 J_x \AC{J_x} \beta_x \AC{\beta_x}} \gray{+ \frac{3}{2} \AC{J_x}^2 \AC{\beta_x}^2} \\
     &\tmplineintro  \colorjxy{- 6 J_x J_y \beta_x \beta_y} \colorjx{- 6 J_x \AC{J_y} \beta_x \AC{\beta_y}} \colorjy{- 6 \AC{J_x} J_y \AC{\beta_x} \beta_y} \gray{- 6 \AC{J_x} \AC{J_y} \AC{\beta_x} \AC{\beta_y}} \\
    &\tmplineintro  \colorjy{+ \frac{3}{2} J_y^2 \beta_y^2 + 6 J_y \AC{J_y} \beta_y \AC{\beta_y}} \gray{+ \frac{3}{2} \AC{J_y}^2 \AC{\beta_y}^2}
    \Bigg) 
\end{split}
\end{align}

During amplitude detuning measurements, we will change the amplitude of the AC-Dipole and hence $\AC{J_{x,y}}$, 
yet we are nor measuring the change in the driven tune - which is fixed by the AC-Dipole frequency - but the change in the natural tune.
While this tune would vanish during a perfectly adiabatically ramp of the AC-Dipole, the presence of non-linearities and non-zero chromaticity
degrades this adiabaticity~\cite{TomasAdiabaticityRampingProcess2005} and the natural tune will be visible in the measured spectrum.
As its amplitude is much smaller than driven tune's, it can get overshadowed by resonances - and sometimes even noise, which makes amplitude detuning
measurements very challenging.

Using~\cref{eq:tuneChangeExpectation}, the change of the natural tunes from normal octupoles under the influence of an AC-Dipole are given as
\begin{align}
\label{eq:deltaQxAC}
\begin{split}
    \Delta Q_x^{\AC{N_4}} 
    \quad &\overeq{\cref{eq:tuneChangeExpectation}} \quad
    \frac{1}{2\pi} \int_L \frac{\partial \langle \AC{N_4} \rangle }{\partial J_x} ds \\
    \quad &\overeq{\cref{eq:n40ExpAC}} \quad
    \frac{1}{16\pi} K_4L \of{\beta^2_xJ_x - 2 \beta_x \beta_y J_y + 2 \beta_x \AC{\beta_x} \AC{J_x} - 2 \beta_x \AC{\beta_y} \AC{J_y} }
\end{split}
\end{align}
\begin{align}
\label{eq:deltaQyAC}
\begin{split}
    \Delta Q_y^{\AC{N_4}} 
    \quad &\overeq{\cref{eq:tuneChangeExpectation}} \quad
    \frac{1}{2\pi} \int_L \frac{\partial \langle \AC{N_4} \rangle }{\partial J_y} ds \\
    \quad &\overeq{\cref{eq:n40ExpAC}} \quad
     \frac{1}{16\pi} K_4L \of{\beta^2_yJ_y - 2 \beta_x \beta_y J_x - 2 \beta_y \AC{\beta_x} \AC{J_x} + 2 \beta_y \AC{\beta_y} \AC{J_y}}
\end{split}
\end{align}

Under the assumption that the AC-Dipole does not affect the $\beta$-function, that is $\AC{\beta_z} \approx \beta_z$, 
the first order detuning with the action $2\AC{J_z}$ is approximately:
\begin{align}
    \frac{\partial Q_x^{\AC{N_4}}}{\partial (2\AC{J_x})} 
    &= \nosign \frac{K_4L}{16\pi} \, \beta_x \AC{\beta_x} 
    \approx \nosign \frac{K_4L}{16\pi} \, \beta^2_x 
    \quad \overeq{\cref{eq:ampdetDirectX}} \quad
     2 \cdot \frac{\partial Q_x^{N_4}}{\partial (2J_x)} 
    \label{eq:ampdetDirectXAC} \\
    \frac{\partial Q_x^{\AC{N_4}}}{\partial (2\AC{J_y})} 
    &= - \frac{K_4L}{16\pi} \, \beta_x \AC{\beta_y} 
    \approx - \frac{K_4L}{16\pi} \, \beta_x \beta_y 
    \quad \overeq{\cref{eq:ampdetCross}} \quad
    \frac{\partial Q_x^{N_4}}{\partial (2J_y)}
    \label{eq:ampdetCrossXAC} \\
    \frac{\partial Q_y^{\AC{N_4}}}{\partial (2\AC{J_x})} 
    &= - \frac{K_4L}{16\pi} \, \AC{\beta_x} \beta_y 
    \approx - \frac{K_4L}{16\pi} \, \beta_x \beta_y 
    \quad \overeq{\cref{eq:ampdetCross}} \quad
    \frac{\partial Q_y^{N_4}}{\partial (2J_x)}
    \label{eq:ampdetCrossYAC} \\
    \frac{\partial Q_y^{\AC{N_4}}}{\partial (2\AC{J_y})}
    &= \nosign \frac{K_4L}{16\pi} \, \beta_y \AC{\beta_y} 
    \approx \nosign \frac{K_4L}{16\pi} \, \beta^2_y 
    \quad \overeq{\cref{eq:ampdetDirectY}} \quad
    2 \cdot \frac{\partial Q_y^{N_4}}{\partial (2J_y)}
    \label{eq:ampdetDirectYAC}
\end{align}
When performing amplitude detuning measurements with the AC-Dipole, 
the detuning of the direct terms will be twice as big as the detuning for free motion. 

\subsection{From Normal Dodecapoles} %--------------------------------------------------------------
\label{sec:AmpDetNormalDodecapoles}
Following the same procedure as in \cref{sec:AmpDetNormalOctupoles} for the normal dodecapole Hamiltonian $N_6$ in \cref{eq:n6O}, one arrives at:
\begin{equation}
    \label{eq:n60CS}
    N_6 = \frac{8}{6!} K_6 \left(J_x^3\beta_x^3 \cos^6{\phi_x} 
    - 15J_x^2 J_y\beta_x^2\beta_y \cos^4{\phi_x}\cos^2{\phi_y} 
    + 15J_x J_y^2\beta_x\beta_y^2 \cos^2{\phi_x}\cos^4{\phi_y} 
    - J_y^3\beta_y^3 \cos^6{\phi_y}\right)
\end{equation}
\begin{equation}
    \label{eq:n60ExpCalc}
    \langle N_6 \rangle = 
    \frac{1}{9} \cdot \frac{1}{32} \; K_6 \left(J_x^3\beta_x^3 
    - 9J_x^2 J_y\beta_x^2\beta_y 
    + 9J_x J_y^2\beta_x\beta_y^2 
    - J_y^3\beta_y^3\right)
\end{equation}
\begin{equation}
    \label{eq:tuneChangeN6}
    \Delta Q_{x,y}^{N_6} = \frac{1}{2\pi} \int_L \frac{\partial \langle N_6 \rangle }{\partial J_{x,y}} ds =
    \frac{1}{9} \cdot \frac{1}{64\pi} \; K_6L
    \frac{\partial}{\partial J_{x,y}}
    \left(J_x^3\beta_x^3 
    - 9J_x^2 J_y\beta_x^2\beta_y 
    + 9J_x J_y^2\beta_x\beta_y^2 
    - J_y^3\beta_y^3\right)
\end{equation}
\begin{equation}
    \label{eq:deltaQxN6}
    \Delta Q_x^{N_6} = 
    \frac{1}{192\pi} \; K_6L
    \left(J_x^2\beta_x^3 
    - 6J_x J_y\beta_x^2\beta_y 
    + 3 J_y^2\beta_x\beta_y^2 
    \right)
\end{equation}
\begin{equation}
    \label{eq:deltaQyN6}
    \Delta Q_y^{N_6} = 
    \frac{1}{192\pi} \; K_6L
    \left(
    - 3J_x^2\beta_x^2\beta_y 
    + 6J_x J_y\beta_x\beta_y^2 
    - J_y^2\beta_y^3\right)
\end{equation}

The first order detuning terms still depend on 
$J_z$ and are hence very small near the beam center.
The second order detuning terms are the ones being independent of action strength:
\begin{equation}
    \label{eq:ampdetN6QxdJxdJx}
    \frac{\partial^2 Q_x^{N_6}}{\partial (2J_x)^2} =  \frac{K_6L}{384\pi} \, \beta^3_x
\end{equation}
\begin{equation}
    \label{eq:ampdetN6QxdJxdJy}
    \frac{\partial^2 Q_x^{N_6}}{\partial (2J_x) \partial(2J_y)} 
    \overset{\cref{eq:tuneChangeN6}}{=} 
     \frac{\partial^2 Q_y^{N_6}}{\partial (2J_x)^2} 
     =
     - \frac{K_6L}{128\pi} \, \beta^2_x \beta_y
\end{equation}
\begin{equation}
    \label{eq:ampdetN6QxdJydJy}
    \frac{\partial^2 Q_x^{N_6}}{\partial (2J_y)^2} 
    \overset{\cref{eq:tuneChangeN6}}{=} 
     \frac{\partial^2 Q_y^{N_6}}{\partial (2J_x) \partial (2J_y)} 
    =  \frac{K_6L}{128\pi} \, \beta_x\beta_y^2
\end{equation}
\begin{equation}
    \label{eq:ampdetN6QydJydJy}
    \frac{\partial^2 Q_y^{N_6}}{\partial (2J_y)^2} =  - \frac{K_6L}{384\pi} \, \beta^3_y
\end{equation}

\subsubsection{With AC-Dipole} %--------------------------------------------------------------------
\label{sec:AmpDetDecapolesACDipole}
As in \cref{sec:AmpDetOctupolesACDipole} we can get the expectation value of $N_6$ under the 
influence of driven oscillation from an AC-Dipole (following the same color scheme as in \cref{eq:n40ExpAC}):
\renewcommand{\tmplineintro}{\phantom{\quad = \quad \frac{1}{6!} K_6 \quad}}
\newcommand{\tmplineintrotwo}{\tmplineintro \phantom{- 15 \quad}}
\begin{align}
\label{eq:n60ExpAC}
\begin{split}
    &\langle \AC{N_6} \rangle 
    \quad \overeq{\cref{eq:n6O}}\quad
    \langle \frac{1}{6!} K_6 \of{x^6 - 15 x^4 y^2 + 15 x^2 y^4 - y^6} \rangle \\
    &\quad \overeq{z \mapsto z + \AC{z}} \quad
    \frac{1}{6!} K_6 \Big( \\
    &\lrangle{x^6} \gray{ + 6 \lrangle{x^5} \lrangle{\AC{x}}} + 15 \lrangle{x^4}\lrangle{\AC{x}^2} \gray{ + 20 \lrangle{x^3} \lrangle{\AC{x}^3}} + 15 \lrangle{x^2}\lrangle{\AC{x}^4} \gray{ + 6 \lrangle{x} \lrangle{\AC{x}^5}} + \lrangle{\AC{x}^6} \\
    &- 15 \left(\lrangle{x^4} \gray{ + 4 \lrangle{x^3} \lrangle{\AC{x}}} + 6 \lrangle{x^2}\lrangle{\AC{x}^2} \gray{+ 4 \lrangle{x}\lrangle{\AC{x}^3}} + \lrangle{\AC{x}^4} \right) \left( \lrangle{y^2} \gray{+ 2 \lrangle{y} \lrangle{\AC{y}}} + \lrangle{\AC{y}^2} \right) \\
    &+ 15 \left(\lrangle{x^2} \gray{+ 2 \lrangle{x} \lrangle{\AC{x}}} + \lrangle{\AC{x}^2} \right) \left(\lrangle{y^4} \gray{ + 4 \lrangle{y^3} \lrangle{\AC{y}}} + 6 \lrangle{y^2}\lrangle{\AC{y}^2} \gray{+ 4 \lrangle{y}\lrangle{\AC{y}^3}} + \lrangle{\AC{y}^4} \right) \\
    &- \lrangle{y^6} \gray{ - 6 \lrangle{y^5} \lrangle{\AC{y}}} - 15 \lrangle{y^4}\lrangle{\AC{y}^2} \gray{ - 20 \lrangle{y^3} \lrangle{\AC{y}^3}} - 15 \lrangle{y^2}\lrangle{\AC{y}^4} \gray{ - 6 \lrangle{y} \lrangle{\AC{y}^5}} - \lrangle{\AC{y}^6} \Big)\\
    &\quad \overeq{\cref{eq:CosOddExp}} \quad
    \frac{1}{6!} K_6 \Big(
    \lrangle{x^6} + 15 \lrangle{x^4}\lrangle{\AC{x}^2} + 15 \lrangle{x^2}\lrangle{\AC{x}^4} + \lrangle{\AC{x}^6} \\
    &\tmplineintro - 15 \Big(
        \lrangle{x^4} \lrangle{y^2} + 6 \lrangle{x^2}\lrangle{\AC{x}^2} \lrangle{y^2} + \lrangle{\AC{x}^4} \lrangle{y^2}  \\
    &\tmplineintrotwo  + \lrangle{x^4} \lrangle{\AC{y}^2} + 6 \lrangle{x^2}\lrangle{\AC{x}^2} \lrangle{\AC{y}^2}+ \lrangle{\AC{x}^4} \lrangle{\AC{y}^2}
    \Big) \\
    &\tmplineintro + 15 \Big( \lrangle{x^2} \lrangle{y^4} + 6 \lrangle{x^2}\lrangle{y^2}\lrangle{\AC{y}^2} + \lrangle{x^2} \lrangle{\AC{y}^4} \\
    &\tmplineintrotwo + \lrangle{\AC{x}^2} \lrangle{y^4} + 6 \lrangle{\AC{x}^2} \lrangle{y^2}\lrangle{\AC{y}^2} + \lrangle{\AC{x}^2} \lrangle{\AC{y}^4} 
    \Big) \\
    &\tmplineintro - \lrangle{y^6} - 15 \lrangle{y^4}\lrangle{\AC{y}^2} - 15 \lrangle{y^2}\lrangle{\AC{y}^4} - \lrangle{\AC{y}^6} \Big)\\
    &\quad \overeq{\substack{\cref{eq:ExpCos} \\ \cref{eq:ActionAngleFreeAC}}} \quad
    \frac{1}{6!} K_6 \Bigg(
    \colorjx{\nosign \frac{5}{2} J_x^3 \beta_x^3 + \frac{45}{2} J_x^2 \AC{J_x} \beta_x^2 \AC{\beta_x} + \frac{45}{2} J_x \AC{J_x}^2 \beta_x \AC{\beta_x}^2} \gray{+ \frac{5}{2} \AC{J_x}^3 \AC{\beta_x}^3}\\
    &\tmplineintro \colorjxy{- \frac{45}{2} J_x^2 J_y \beta_x^2 \beta_y} \colorjxy{- 90 J_x \AC{J_x} J_y \beta_x \AC{\beta_x} \beta_y} \colorjy{- \frac{45}{2} \AC{J_x}^2 J_y \AC{\beta_x}^2 \beta_y}  \\
    &\tmplineintro \colorjx{- \frac{45}{2} J_x^2 \AC{J_y} \beta_x^2 \AC{\beta_y}} \colorjx{- 90 J_x \AC{J_x} \AC{J_y} \beta_x \AC{\beta_x} \AC{\beta_y}}  \gray{- \frac{45}{2}  \AC{J_x}^2 \AC{J_y} \AC{\beta_x}^2 \AC{\beta_y}}\\ 
    &\tmplineintro \colorjxy{+ \frac{45}{2} J_x J_y^2 \beta_x \beta_y^2} \colorjxy{+ 90 J_x J_y \AC{J_y} \beta_x \beta_y \AC{\beta_y}} \colorjx{+ \frac{45}{2} J_x \AC{J_y}^2 \beta_x \AC{\beta_y}^2} \\
    &\tmplineintro \colorjy{+ \frac{45}{2} \AC{J_x} J_y^2 \AC{\beta_x} \beta_y^2 + 90 \AC{J_x} J_y \AC{J_y} \AC{\beta_x} \beta_y \AC{\beta_y}}  \gray{+ \frac{45}{2}  \AC{J_x} \AC{J_y}^2 \AC{\beta_x} \AC{\beta_y}^2}\\
    &\tmplineintro \colorjy{- \frac{5}{2} J_y^3 \beta_y^3 - \frac{45}{2} J_y^2 \AC{J_y} \beta_y^2 \AC{\beta_y} - \frac{45}{2} J_y \AC{J_y}^2 \beta_y \AC{\beta_y}^2} \gray{ - \frac{5}{2} \AC{J_y}^3 \AC{\beta_y}^3}\Bigg) \\
\end{split}
\end{align}

\clearpage
The tune change from a dodecapole under driven oscillation is therefore (terms are kept in the same order and lines as in~\cref{eq:n60ExpAC} for consistency):
\renewcommand{\tmplineintro}{\phantom{ = \quad \frac{1}{192\pi} K_6L \big(}}
\begin{align}
\label{eq:deltaQxACn6O}
\begin{split}
    \Delta Q_x^{\AC{N_6}} 
    \quad &\overeq{\cref{eq:tuneChangeExpectation}} \quad
    \frac{1}{2\pi} \int_L \frac{\partial \langle \AC{N_6} \rangle }{\partial J_x} ds \\
    \quad &\overeq{\cref{eq:n60ExpAC}} \quad
    \frac{1}{192\pi} K_6L \big( 
    \gray{\beta^3_x J_x^2} \colorjx{\gray{+ 6 \beta_x \AC{\beta_x} J_x \AC{J_x}}} \colorjxx{+ 3 \beta_x \AC{\beta_x}^2 \AC{J_x}^2} \\ 
    & \tmplineintro \gray{- 6 \beta_x^2 \beta_y J_x J_y} \colorjx{\gray{- 12 \beta_x \AC{\beta_x} \beta_y \AC{J_x} J_y}}  \\
    & \tmplineintro \colorjy{\gray{- 6 \beta_x^2 \AC{\beta_y} J_x \AC{J_y}}} \colorjxy{- 12 \beta_x \AC{\beta_x} \AC{\beta_y} \AC{J_x} \AC{J_y}} \\
    & \tmplineintro \gray{+ 3 \beta_x \beta_y^2 J_y^2} \colorjy{\gray{+ 12 \beta_x \beta_y \AC{\beta_y}  J_y \AC{J_y}}} \colorjyy{+ 3 \beta_x \AC{\beta_y}^2 \AC{J_y}^2} 
    \big)
\end{split}
\end{align}
\begin{align}
\label{eq:deltaQyACn6O}
\begin{split}
    \Delta Q_y^{\AC{N_6}} 
    \quad &\overeq{\cref{eq:tuneChangeExpectation}} \quad
    \frac{1}{2\pi} \int_L \frac{\partial \langle \AC{N_6} \rangle }{\partial J_y} ds \\
    \quad &\overeq{\cref{eq:n60ExpAC}} \quad
     \frac{1}{192\pi} K_6L \big( 
         \gray{- 3 \beta_x^2 \beta_y J_x^2} \colorjx{\gray{- 12 \beta_x \AC{\beta_x} \beta_y J_x \AC{J_x}}} \colorjxx{-3 \AC{\beta_x}^2 \beta_y \AC{J_x}^2} \\ 
         & \tmplineintro \gray{+6 \beta_x \beta_y^2 J_x J_y} \colorjy{\gray{+ 12 \beta_x \beta_y \AC{\beta_y} J_x \AC{J_y}}} \\ 
         & \tmplineintro \colorjx{\gray{+ 6 \AC{\beta_x} \beta_y^2 \AC{J_x} J_y}} \colorjxy{+12 \AC{\beta_x} \beta_y \AC{\beta_y} \AC{J_x} \AC{J_y}} \\ 
         & \tmplineintro \gray{- \beta_y^3 J_y^2} \colorjy{\gray{- 6 \beta_y^2 \AC{\beta_y} J_y \AC{J_y}}} \colorjyy{- 3 \beta_y \AC{\beta_y}^2 \AC{J_y}^2}
     \big) 
\end{split}
\end{align}
Terms that are of no further interest as they do not contain $\AC{J_x}^2$, $\AC{J_y}^2$ or $\AC{J_x}\AC{J_y}$are greyed out; terms depending only on $\AC{J_x}$ are underlined blue,
on $\AC{J_y}$ are underlined red.
All terms of the first derivatives in $\AC{J_x}$ or $\AC{J_y}$ of \cref{eq:deltaQxACn6O,eq:deltaQyACn6O} will still depend 
on some $J_z$ or $\AC{J_z}$ and are therefore, again, very small.
Only the second order detuning with the action $2\AC{J_z}$ will be constant:
\begin{align}
    \label{eq:ampdetN6QxdJxdJxAC}
    &\frac{\partial^2 Q_x^{\AC{N_6}}}{\partial (2\AC{J_x})^2} 
    &&\overeq{\cref{eq:deltaQxACn6O}} \quad 
    \nosign \frac{K_6L}{128\pi} \, \beta_x \AC{\beta_x}^2 
    &&\approx
    \nosign \frac{K_6L}{128\pi} \, \beta_x^3 
    &&\overeq{\cref{eq:ampdetN6QxdJxdJx}} \quad
    3 \cdot \frac{\partial^2 Q_x^{N_6}}{\partial (2J_x)^2} 
    \\
    \label{eq:ampdetN6QxdJxdJyAC}
    &\frac{\partial^2 Q_x^{\AC{N_6}}}{\partial (2\AC{J_x}) \partial(2\AC{J_y})} 
    &&\overeq{\cref{eq:deltaQxACn6O}} \quad
    - \frac{K_6L}{64\pi} \, \beta_x \AC{\beta_x} \AC{\beta_y}
    &&\approx 
    - \frac{K_6L}{64\pi} \, \beta_x^2 \beta_y
    &&\overeq{\cref{eq:ampdetN6QxdJxdJy}} \quad
    2 \cdot \frac{\partial^2 Q_x^{N_6}}{\partial (2J_x) \partial(2J_y)} 
    \\ 
    \label{eq:ampdetN6QxdJydJyAC}
    &\frac{\partial^2 Q_x^{\AC{N_6}}}{\partial (2\AC{J_y})^2} 
    &&\overeq{\cref{eq:deltaQxACn6O}}  \quad
    \nosign \frac{K_6L}{128\pi} \, \beta_x \AC{\beta_y}^2
    &&\approx 
    \nosign \frac{K_6L}{128\pi} \, \beta_x \beta_y^2
    &&\overeq{\cref{eq:ampdetN6QxdJydJy}} \quad
    \phantom{2 \cdot} \frac{\partial^2 Q_x^{N_6}}{\partial (2J_y)^2} 
    \\
    \label{eq:ampdetN6QydJxdJxAC}
    &\frac{\partial^2 Q_y^{\AC{N_6}}}{\partial (2\AC{J_x})^2} 
    &&\overeq{\cref{eq:deltaQyACn6O}}  \quad
    - \frac{K_6L}{128\pi} \, \AC{\beta_x}^2 \beta_y 
    &&\approx 
    - \frac{K_6L}{128\pi} \, \beta_x^2 \beta_y
    &&\overeq{\cref{eq:ampdetN6QxdJxdJy}} \quad
    \phantom{2 \cdot}\frac{\partial^2 Q_y^{N_6}}{\partial (2J_x)^2} 
    \\
    \label{eq:ampdetN6QydJxdJyAC}
    &\frac{\partial^2 Q_y^{\AC{N_6}}}{\partial (2\AC{J_x}) \partial (2\AC{J_y})} 
    &&\overeq{\cref{eq:deltaQyACn6O}}  \quad
    \nosign \frac{K_6L}{64\pi} \,  \AC{\beta_x} \beta_y \AC{\beta_y} 
    &&\approx 
    \nosign \frac{K_6L}{64\pi} \,  \beta_x \beta_y^2 
    &&\overeq{\cref{eq:ampdetN6QxdJydJy}} \quad
    2 \cdot \frac{\partial^2 Q_y^{N_6}}{\partial (2J_x) \partial (2J_y)} 
    \\
    \label{eq:ampdetN6QydJydJyAC}
    &\frac{\partial^2 Q_y^{\AC{N_6}}}{\partial (2\AC{J_y})^2} 
    &&\overeq{\cref{eq:deltaQyACn6O}} \quad 
    - \frac{K_6L}{128\pi} \, \beta_y \AC{\beta_y}^2
    &&\approx  
    - \frac{K_6L}{128\pi} \, \beta_y^3
    &&\overeq{\cref{eq:ampdetN6QydJydJy}} \quad
    3 \cdot \frac{\partial^2 Q_y^{N_6}}{\partial (2J_y)^2} 
\end{align}
In conclusion, the single-plane cross-terms of second order amplitude detuning are not influenced by the AC-Dipole, 
yet they loose their equality with the diagonal detuning, which are under driven motion is twice as high as under free motion.
The direct terms on the other hand will yield three times the free-motion detuning during AC-Dipole measurements,
in agreement with the results from \cite{WhiteDirectAmplitudeDetuning2013}, in which the diagonal terms were not investigated.

\subsection{General Dependencies on Field Order} %--------------------------------------------------
\label{sec:AmpDetGeneralDependenciesOnFieldOrder}
While in \cref{sec:AmpDetNormalOctupoles,sec:AmpDetNormalDodecapoles} the first order contributions 
to first and second order amplitude detuning is presented, 
second, third and fourth order contributions can be identified as well
by creating combinations whose terms do not vanish during the derivative steps as outlined above.
Some of them are given in \cref{tab:ampDetDependence}.

\begin{table}[!h]
    \centering
    \begin{tabular}{c|c}
         Order & Source \\
         $\frac{\partial Q_z}{\partial (2J_z)}$ & $K_3^2$, $K_4$ \\
         $\frac{\partial^2 Q_z}{\partial (2J_z)^2}$ & $K_3^4$, $K_3^2K_4$, $K_4^2$, $K_3K_5$, $K_6$ 
    \end{tabular}
    \caption{Contributions to amplitude detuning order from field strength combinations~\cite{Bazzani10EffectsSystematic1994}.}
    \label{tab:ampDetDependence}
\end{table}
\section{Chromaticity}
\label{sec:Chromaticity}
Chromaticity is the change of tune with change of the particles energy or momentum.
With the relative momentum change $\delta = \frac{\delta p}{p}$ we can expand the tune $Q_z$ around its design value at the momentum $p$
\begin{equation}
    Q_z (\delta) = Q_z + Q'_z \delta + \frac{1}{2!} Q''_z \delta^2 + \frac{1}{3!} Q'''_z \delta^3 + \frac{1}{4!} Q''''_z \delta^4 + h.o.t
\end{equation}
The Q-n-prime coefficients are called the $n^{th}$ order chromaticity respectively, and can be calculated by taking the $n^{th}$ derivative
with respect to $\delta$
\begin{equation}
    Q^{n\prime} = \frac{\partial^n Q_z}{\pardelta^n} \;.
\end{equation}

The contribution from a magnet to chromaticity can be calculated in a similar manner as Amplitude Detuning in \cref{sec:AmplitudeDetuning},
as the momentum change $\delta$ introduces a change in orbit, we can expand the Hamiltonian of the magnetic field 
as has been done for the feed-down in \cref{sec:FeedDown}.
The Dispersion coefficient $D_z$ determines the change in orbit:
\begin{equation}
    \Delta z = D_z \delta \;.
\end{equation}
The contribution of one magnet to the change in tune via its Hamiltonian $H$ is therefore
\begin{equation}
    \label{eq:tuneChangeChroma}
    \frac{\partial^n \Delta Q_z}{\pardelta^n} \quad \overeq{\cref{eq:tuneChangeExpectation}} \quad
    \frac{1}{2\pi} \int_L \frac{\partial^{n+1} H(x + D_x \delta, \; y + D_y \delta)}{\pardelta^n \; \partial J_z} ds \;.
\end{equation}

For easier reading, some terms in the equations are greyed out, either because they contain odd cosine and vanish due to \cref{eq:CosOddExp}
or because they do not depend on $\delta$ and hence play no role for chromaticity.
The Taylor series are also only expanded up to the last term still dependent on $x$ or $y$ as any additional
term would vanish under the derivative with respect to the action in a later step.

\subsection{From Dipoles and Quadrupoles}
In first order, the contribution from dipoles equals zero because of the derivation of $J_z$ followed by $\delta$,
while for quadrupoles the terms vanish according to \cref{eq:CosOddExp} as there are only odd powers of cosine present in the expanded terms of
$N_2\of{x+D_x\delta, y+D_y\delta}$ and $S_2\of{x+D_x\delta, y+D_y\delta}$, which do not vanish on performing the derivative with respect to $\delta$.

\subsection{From Sextupoles} %---------------------------------------------------------------------- 
Sextupoles contribute to first order chromaticity as shown in the following. 
Any further derivative with respect to $\delta$ vanish, and hence in first order sextupoles do not contribute to higher order chromaticity.

\subsubsection{Normal Sextupole}
\begin{align}
\label{eq:n3OmomentumExpectation}
\begin{split}
    \langle N_3\of{x+D_x\delta, y+D_y\delta} \rangle 
    \quad &\myover{\substack{\cref{eq:n3O} \\ + Taylor}}{\approx} \quad
    \langle \gray{\frac{1}{3!} K_3 \of{x^3 - 3 x y^2}} + \frac{1}{2}K_3\of{x^2 - y^2}D_x\delta \gray{- K_3 x y D_y\delta} \\
    &\qquad \gray{+ \frac{1}{2} K_3 x \of{D_x^2 - D_y^2} \delta^2 + K_3 y D_x D_y \delta^2} \rangle\\
    \quad &\overeq{\cref{eq:CosOddExp}} \quad
    \frac{1}{2} K_3 \of{2J_x\beta_x\langle \cos^2 \phi_x \rangle - 2J_y\beta_y \langle \cos^2 \phi_y \rangle } D_x \delta  \\
    \quad &\overeq{\cref{eq:ExpCos}} \quad
    \frac{1}{2} K_3 \of{J_x\beta_x - J_y\beta_y }D_x \delta 
\end{split}
\end{align}

\begin{align}
    \frac{\partial \Delta Q_x^{N_3}}{\pardelta}
    \quad &\overeq{\cref{eq:tuneChangeChroma}} \quad
    \frac{1}{2\pi} \int_L \frac{\partial^2 \langle N_3\of{x+D_x\delta, y+D_y\delta} \rangle }{\pardelta \; \partial J_x} ds
    \quad \myover{\cref{eq:n3OmomentumExpectation}}{\approx} \quad
    \nosign \frac{1}{4\pi}K_3L\beta_x D_x
    \label{eq:chromaXN3}
    \\
    \frac{\partial \Delta Q_y^{N_3}}{\pardelta} 
    \quad &\overeq{\cref{eq:tuneChangeChroma}} \quad
    \frac{1}{2\pi} \int_L \frac{\partial^2 \langle N_3\of{x+D_x\delta, y+D_y\delta} \rangle }{\pardelta \; \partial J_y} ds 
    \quad \myover{\cref{eq:n3OmomentumExpectation}}{\approx} \quad
    - \frac{1}{4\pi}K_3L\beta_y D_x 
    \label{eq:chromaYN3}
\end{align}

\subsubsection{Skew Sextupole}
\begin{align}
\label{eq:s3OmomentumExpectation}
\begin{split}
    \langle S_3\of{x+D_x\delta, y+D_y\delta} \rangle 
    \quad &\myover{\substack{\cref{eq:s3O} \\ + Taylor}}{\approx} \quad
    \langle \gray{ -\frac{1}{3!} J_3 \of{3x^2y - y^3} - J_3 x y D_x\delta} - \frac{1}{2} J_3\of{x^2 - y^2}D_y\delta \\
    &\qquad \gray{- \frac{1}{2} J_3 x \of{Dx^2 - D_y^2} \delta^2 - J_3 y D_x D_y \delta^2} \rangle\\
    \quad &\overeq{\cref{eq:CosOddExp}} \quad
    -\frac{1}{2} J_3 \of{2J_x\beta_x\langle \cos^2 \phi_x \rangle - 2J_y\beta_y \langle \cos^2 \phi_y \rangle } D_y \delta \\
    \quad &\overeq{\cref{eq:ExpCos}} \quad
    -\frac{1}{2} J_3  \of{J_x\beta_x - J_y\beta_y }D_y \delta
\end{split}
\end{align}

\begin{align}
    \frac{\partial \Delta Q_x^{S_3}}{\pardelta}
    \quad &\overeq{\cref{eq:tuneChangeChroma}} \quad
    \frac{1}{2\pi} \int_L \frac{\partial^2 \langle S_3\of{x+D_x\delta, y+D_y\delta} \rangle }{\pardelta \; \partial J_x} ds
    \quad \myover{\cref{eq:s3OmomentumExpectation}}{\approx} \quad
    -\frac{1}{4\pi}J_3L\beta_x D_y 
    \label{eq:chromaXS3} \\
    \frac{\partial \Delta Q_y^{S_3}}{\pardelta} 
    \quad &\overeq{\cref{eq:tuneChangeChroma}} \quad
    \frac{1}{2\pi} \int_L \frac{\partial^2 \langle S_3\of{x+D_x\delta, y+D_y\delta} \rangle }{\pardelta \; \partial J_y} ds 
    \quad \myover{\cref{eq:s3OmomentumExpectation}}{\approx} \quad
    \nosign \frac{1}{4\pi}J_3L\beta_y D_y
    \label{eq:chromaYS3}
\end{align}

\subsection{From Octupoles} %-----------------------------------------------------------------------
\subsubsection{Normal Octupole}
\begin{align}
\label{eq:n4OmomentumExpectation}
\begin{split}
    \langle N_4\of{x+D_x\delta, y+D_y\delta} \rangle 
    \quad &\myover{\substack{\cref{eq:n4O} \\ + Taylor}}{\approx} \quad
    \langle 
        \frac{1}{4!} K_4 \of{x^4 - 6 x^2 y^2 + y^4} \\
        &\quad\quad \gray{+ \frac{1}{3!}K_4\of{x^3 - 3 y x^2}D_x\delta - \frac{1}{3!}K_4 \of{3x^2y-y^3} D_y\delta} \\
        &\quad\quad + \frac{1}{4}K_4 \of{x^2-y^2} \of{D_x^2-D_y^2}\delta^2 \gray{- K_4 x y D_x D_y \delta^2}  \\
        &\quad\quad \gray{+ \frac{1}{6} K_4 x \of{D_x^3 - 3 D_x D_y^2} \delta^3 - \frac{1}{6} K_4 y \of{3 D_x^2 D_y - D_y^3} \delta^3}
    \rangle\\
    \quad &\overeq{\cref{eq:CosOddExp}} \quad
    \gray{\langle  \frac{1}{4!} K_4 \of{x^4 - 6 x^2 y^2 + y^4} \rangle} \\
    &\quad\quad +\frac{1}{4} K_4 \of{2J_x\beta_x\langle \cos^2 \phi_x \rangle - 2J_y\beta_y \langle \cos^2 \phi_y \rangle } \of{D_x^2-D_y^2}\delta^2  \\
    \quad &\overeq{\cref{eq:ExpCos}} \quad
    \gray{\langle  \frac{1}{4!} K_4 \of{x^4 - 6 x^2 y^2 + y^4} \rangle} \\
    &\quad\quad +\frac{1}{4} K_4 \of{J_x\beta_x - J_y\beta_y} \of{D_x^2-D_y^2}\delta^2  \\
\end{split}
\end{align}
$Q'$ on will still depend on $\delta$ and will therefore be very small.
Hence, normal octupoles contribute mainly to second order chromaticity.

\begin{align}
    \frac{\partial^2 \Delta Q_x^{N_4}}{\pardelta^2}
    \quad &\overeq{\cref{eq:tuneChangeChroma}} \quad
    \frac{1}{2\pi} \int_L \frac{\partial^3 \langle N_4\of{x+D_x\delta, y+D_y\delta} \rangle }{\pardelta^2 \; \partial J_x} ds
    \quad \myover{\cref{eq:n4OmomentumExpectation}}{\approx} \quad
    \nosign \frac{1}{4\pi}K_4L\beta_x \of{D_x^2-D_y^2}
    \label{eq:secondchromaXN4}
    \\
    \frac{\partial^2 \Delta Q_y^{N_4}}{\pardelta^2} 
    \quad &\overeq{\cref{eq:tuneChangeChroma}} \quad
    \frac{1}{2\pi} \int_L \frac{\partial^3 \langle N_4\of{x+D_x\delta, y+D_y\delta} \rangle }{\pardelta^2 \; \partial J_y} ds 
    \quad \myover{\cref{eq:n4OmomentumExpectation}}{\approx} \quad
    - \frac{1}{4\pi}K_4L\beta_y \of{D_x^2-D_y^2}
    \label{eq:secondchromaYN4}
\end{align}

\subsubsection{Skew Octupole}
\begin{align}
\label{eq:s4OmomentumExpectation}
\begin{split}
    \langle S_4\of{x+D_x\delta, y+D_y\delta} \rangle 
    \quad &\myover{\substack{\cref{eq:s4O} \\ + Taylor}}{\approx} \quad
    \langle 
        \gray{- \frac{1}{4!} J_4 \of{4 x^3 y - 4 x y^3}} \\
        &\quad\quad \gray{ - \frac{1}{3!}J_4\of{3 x^2 y - y^3}D_x\delta - \frac{1}{3!}J_4 \of{x^3 - 3 x^2 y} D_y\delta }\\
        &\quad\quad \gray{- \frac{1}{2} J_4 x y  \of{D_x^2-D_y^2}\delta^2} - \frac{1}{2} J_4 \of{x^2-y^2} D_x D_y \delta^2  \\
        &\quad\quad \gray{- \frac{1}{6} J_4 y \of{ D_x^3 - 3 D_x D_y^2} \delta^3 - \frac{1}{6} J_4 x \of{3 D_x^2 D_y - D_y^3} \delta^3}
    \rangle\\
    \quad &\overeq{\cref{eq:CosOddExp}} \quad
    -\frac{1}{2} J_4 \of{2J_x\beta_x\langle \cos^2 \phi_x \rangle - 2J_y\beta_y \langle \cos^2 \phi_y \rangle } D_x D_y \delta^2  \\
    \quad &\overeq{\cref{eq:ExpCos}} \quad
    -\frac{1}{2} J_4 \of{J_x\beta_x - J_y\beta_y} D_x D_y \delta^2 \\
\end{split}
\end{align}
As in the case of the normal octupoles, the contribution to $Q'$ scales with $\delta$ and 
therefore also skew octupoles contribute mainly to second order chromaticity.

\begin{align}
    \frac{\partial^2 \Delta Q_x^{S_4}}{\pardelta^2}
    \quad &\overeq{\cref{eq:tuneChangeChroma}} \quad
    \frac{1}{2\pi} \int_L \frac{\partial^3 \langle S_4\of{x+D_x\delta, y+D_y\delta} \rangle }{\pardelta^2 \; \partial J_x} ds
    \quad \myover{\cref{eq:s4OmomentumExpectation}}{\approx} \quad
    - \frac{1}{2\pi}J_4L\beta_x D_x D_y 
    \label{eq:secondchromaXS4}
    \\
    \frac{\partial^2 \Delta Q_y^{S_4}}{\pardelta^2} 
    \quad &\overeq{\cref{eq:tuneChangeChroma}} \quad
    \frac{1}{2\pi} \int_L \frac{\partial^3 \langle S_4\of{x+D_x\delta, y+D_y\delta} \rangle }{\pardelta^2 \; \partial J_y} ds 
    \quad \myover{\cref{eq:s4OmomentumExpectation}}{\approx} \quad
    \nosign \frac{1}{2\pi}J_4L\beta_y D_x D_y
    \label{eq:secondchromaYS4}
\end{align}

\subsection{From Decapoles} %-----------------------------------------------------------------------
\subsubsection{Normal Decapole}
\label{sec:ChromaFromNormalDecapole}

\begin{align}
\label{eq:n5OmomentumExpectation}
\begin{split}
    &\langle N_5\of{x+D_x\delta, y+D_y\delta} \rangle \\
    \quad &\myover{\substack{\cref{eq:n5O} \\ + Taylor}}{\approx} \quad
        \langle 
        \gray{\frac{1}{5!} K_5 \of{x^5 - 10 x^3 y^2 + 5 x y^4}} \\
        &\quad\quad+ \frac{1}{4!}K_5\of{x^4 - 6 x^2 y^2 + y^4}D_x\delta \gray{- \frac{1}{4!}K_5 \of{4 x^3 y - 4 x y^3} D_y\delta} \\
        &\quad\quad\gray{+ \frac{1}{12}K_5 \of{x^3 - 3 x y^2} \of{D_x^2-D_y^2}\delta^2 - \frac{1}{3!} K_5 \of{3 x^2 y - y^3} D_x D_y \delta^2}  \\
        &\quad\quad+ \frac{1}{12} K_5 \of{x^2 - y^2} \of{D_x^3 - 3 D_x D_y^2} \delta^3 \gray{- \frac{1}{6} K_5 x y  \of{3 D_x^2 D_y - D_y^3} \delta^3} \\
        &\quad\quad\gray{+ \frac{1}{24} K_5 x \of{ D_x^4 - 6 D_x^2 D_y^2 + D_y^4} \delta^4 - \frac{1}{6} K_5 y  \of{D_x^3 D_y - D_x D_y^3} \delta^4}
    \rangle\\
    \quad &\overeq{\cref{eq:CosOddExp}} \quad
    \frac{1}{6} K_5 \of{J_x^2\beta_x^2\langle \cos^4 \phi_x \rangle - 6 J_x\beta_x\langle \cos^2 \phi_x \rangle J_y\beta_y \langle \cos^2 \phi_y \rangle + J_y^2\beta_y^2 \langle \cos^4 \phi_y \rangle } D_x\delta  \\
    &\quad\quad +\frac{1}{12} K_5 \of{2J_x\beta_x\langle \cos^2 \phi_x \rangle - 2J_y\beta_y \langle \cos^2 \phi_y \rangle } \of{D_x^3 - 3 D_x D_y^2} \delta^3  \\
    \quad &\overeq{\cref{eq:ExpCos}} \quad
    \frac{1}{16} K_5 \of{J_x^2\beta_x^2 - 4 J_x\beta_x J_y\beta_y  + J_y^2\beta_y^2} D_x\delta  \\
    &\quad\quad +\frac{1}{12} K_5 \of{J_x\beta_x - J_y\beta_y} \of{D_x^3 - 3 D_x D_y^2} \delta^3  \\
\end{split}
\end{align}
\begin{align}
    \begin{split}
        \frac{\partial \Delta Q_x^{N_5}}{\pardelta}
        \quad &\overeq{\cref{eq:tuneChangeChroma}} \quad
        \frac{1}{2\pi} \int_L \frac{\partial^2 \langle N_5\of{x+D_x\delta, y+D_y\delta} \rangle }{\pardelta \; \partial J_x} ds \\
        \quad &\myover{\cref{eq:n5OmomentumExpectation}}{\approx} \quad
        \nosign \frac{1}{16\pi} K_5L \of{J_x\beta_x^2 - 2J_y\beta_x \beta_y} D_x 
        +\frac{1}{8\pi} K_5L \beta_x \of{D_x^3 - 3 D_x D_y^2} \delta^2
        \label{eq:chromaXN5}
    \end{split}
    \\
    \begin{split}
        \frac{\partial \Delta Q_y^{N_5}}{\pardelta} 
        \quad &\overeq{\cref{eq:tuneChangeChroma}} \quad
        \frac{1}{2\pi} \int_L \frac{\partial^2 \langle N_5\of{x+D_x\delta, y+D_y\delta} \rangle }{\pardelta \; \partial J_y} ds \\
        \quad &\myover{\cref{eq:n5OmomentumExpectation}}{\approx} \quad
        -\frac{1}{16\pi} K_5L \of{2J_x\beta_x \beta_y  - J_y\beta_y^2} D_x
         -\frac{1}{8\pi} K_5L  \beta_y \of{D_x^3 - 3 D_x D_y^2} \delta^2
        \label{eq:chromaYN5}
    \end{split}
\end{align}
The first term still depends on $J_x$ and $J_y$, while the second term is proportional to $\delta^2$. 
The contribution of $N_5$ to $Q'$ is therefore very small. 
Similarly, $Q''$ will also still depend on $\delta$ and is hence small.
In conclusion, normal decapoles mainly contribute to $Q'''$:

\begin{align}
    \frac{\partial^3 \Delta Q_x^{N_5}}{\pardelta^3}
    \quad &\overeq{\cref{eq:tuneChangeChroma}} \quad
    \frac{1}{2\pi} \int_L \frac{\partial^4 \langle N_5\of{x+D_x\delta, y+D_y\delta} \rangle }{\pardelta^3 \; \partial J_x} ds
    \quad \myover{\cref{eq:n5OmomentumExpectation}}{\approx} \quad
    \nosign \frac{1}{4\pi} K_5L \beta_x \of{D_x^3 - 3 D_x D_y^2}
    \label{eq:thirdchromaXN5}
    \\
    \frac{\partial^3 \Delta Q_y^{N_5}}{\pardelta^3} 
    \quad &\overeq{\cref{eq:tuneChangeChroma}} \quad
    \frac{1}{2\pi} \int_L \frac{\partial^4 \langle N_5\of{x+D_x\delta, y+D_y\delta} \rangle }{\pardelta^3 \; \partial J_y} ds 
    \quad \myover{\cref{eq:n5OmomentumExpectation}}{\approx} \quad
    -\frac{1}{4\pi} K_5L  \beta_y \of{D_x^3 - 3 D_x D_y^2}
    \label{eq:thirdchromaYN5}
\end{align}

\subsubsection{Skew Decapole}
\begin{align}
\label{eq:s5OmomentumExpectation}
\begin{split}
    &\langle S_5\of{x+D_x\delta, y+D_y\delta} \rangle \\
    \quad &\myover{\substack{\cref{eq:s5O} \\ + Taylor}}{\approx} \quad
        \langle 
        \gray{-\frac{1}{5!} J_5 \of{5x^4 y - 10 x^2 y^3 + y^5}} \\
        &\quad\quad\gray{- \frac{1}{4!}J_5\of{4x^3 y - x y^3}D_x\delta} - \frac{1}{4!}J_5 \of{x^4 - 6 x^2 y^2 + y^4} D_y\delta \\
        &\quad\quad\gray{- \frac{1}{12}J_5 \of{3 x^2 y - y^3} \of{D_x^2-D_y^2}\delta^2 - \frac{1}{3!} J_5 \of{x^3 - 3 x y^2} D_x D_y \delta^2}  \\
        &\quad\quad\gray{- \frac{1}{6} J_5 x y \of{D_x^3 - 3 D_x D_y^2} \delta^3} - \frac{1}{12} J_5 \of{x^2 - y^2} \of{3 D_x^2 D_y - D_y^3} \delta^3 \\
        &\quad\quad\gray{- \frac{1}{24} J_5 y \of{ D_x^4 - 6 D_x^2 D_y^2 + D_y^4} \delta^4 - \frac{1}{6} J_5 x  \of{D_x^3 D_y - D_x D_y^3} \delta^4}
    \rangle\\
    \quad &\overeq{\cref{eq:CosOddExp}} \quad
    -\frac{1}{6} J_5 \of{J_x^2\beta_x^2\langle \cos^4 \phi_x \rangle -J_x\beta_x\langle \cos^2 \phi_x \rangle J_y\beta_y \langle \cos^2 \phi_y \rangle + J_y^2\beta_y^2 \langle \cos^4 \phi_y \rangle } D_y\delta  \\
    &\quad\quad -\frac{1}{12} J_5 \of{2J_x\beta_x\langle \cos^2 \phi_x \rangle - 2J_y\beta_y \langle \cos^2 \phi_y \rangle } \of{3 D_x^2 D_y - D_y^3} \delta^3  \\
    \quad &\overeq{\cref{eq:ExpCos}} \quad
    -\frac{1}{16} J_5 \of{J_x^2\beta_x^2 -4 J_x\beta_x J_y\beta_y  + J_y^2\beta_y^2} D_y\delta  \\
    &\quad\quad -\frac{1}{12} J_5 \of{J_x\beta_x - J_y\beta_y} \of{3 D_x^2 D_y - D_y^3} \delta^3  \\
\end{split}
\end{align}
\begin{align}
    \begin{split}
        \frac{\partial \Delta Q_x^{S_5}}{\pardelta}
        \quad &\overeq{\cref{eq:tuneChangeChroma}} \quad
        \frac{1}{2\pi} \int_L \frac{\partial^2 \langle S_5\of{x+D_x\delta, y+D_y\delta} \rangle }{\pardelta \; \partial J_x} ds \\
        \quad &\myover{\cref{eq:n5OmomentumExpectation}}{\approx} \quad
        -\frac{1}{16\pi} J_5L \of{J_x\beta_x^2 - 2 J_y\beta_x \beta_y} D_y 
        -\frac{1}{8\pi} J_5L \beta_x \of{3 D_x^2 D_y - D_y^3} \delta^2
        \label{eq:chromaXS5}
    \end{split}
    \\
    \begin{split}
        \frac{\partial \Delta Q_y^{S_5}}{\pardelta} 
        \quad &\overeq{\cref{eq:tuneChangeChroma}} \quad
        \frac{1}{2\pi} \int_L \frac{\partial^2 \langle S_5\of{x+D_x\delta, y+D_y\delta} \rangle }{\pardelta \; \partial J_y} ds \\
        \quad &\myover{\cref{eq:n5OmomentumExpectation}}{\approx} \quad
        \nosign \frac{1}{16\pi} J_5L \of{2J_x\beta_x \beta_y  - J_y\beta_y^2} D_y
        +\frac{1}{8\pi} J_5L  \beta_y \of{3 D_x^2 D_y - D_y^3} \delta^2
        \label{eq:chromaYS5}
    \end{split}
\end{align}
With the same arguments as for the normal decapoles in \cref{sec:ChromaFromNormalDecapole} $Q'$ and $Q''$ can therefore be neglected 
and skew decapoles also mainly contribute to $Q'''$:
\begin{align}
    \frac{\partial^3 \Delta Q_x^{S_5}}{\pardelta^3}
    \quad &\overeq{\cref{eq:tuneChangeChroma}} \quad
    \frac{1}{2\pi} \int_L \frac{\partial^4 \langle S_5\of{x+D_x\delta, y+D_y\delta} \rangle }{\pardelta^3 \; \partial J_x} ds
    \quad \myover{\cref{eq:s5OmomentumExpectation}}{\approx} \quad
    - \frac{1}{4\pi} J_5L \beta_x \of{3 D_x^2 D_y - D_y^3}
    \label{eq:thirdchromaXS5}
    \\
    \frac{\partial^3 \Delta Q_y^{S_5}}{\pardelta^3} 
    \quad &\overeq{\cref{eq:tuneChangeChroma}} \quad
    \frac{1}{2\pi} \int_L \frac{\partial^4 \langle S_5\of{x+D_x\delta, y+D_y\delta} \rangle }{\pardelta^3 \; \partial J_y} ds 
    \quad \myover{\cref{eq:s5OmomentumExpectation}}{\approx} \quad
    \nosign \frac{1}{4\pi} J_5L  \beta_y \of{3 D_x^2 D_y - D_y^3}
    \label{eq:thirdchromaYS5}
\end{align}
\pagebreak

\subsection{From Dodecapoles} %--------------------------------------------------------------------- 

\subsubsection{Normal Dodecapole}
\label{sec:ChromaFromNormalDodecapole}

\begin{align}
\label{eq:n6OmomentumExpectation}
\begin{split}
    &\langle N_6\of{x+D_x\delta, y+D_y\delta} \rangle \\
    \quad &\myover{\substack{\cref{eq:n6O} \\ + Taylor}}{\approx} \quad
        \langle 
        \frac{1}{6!} K_6 \of{x^6 - 15 x^4 y^2 + 15 x^2 y^4 - y^6} \\
        &\quad\quad\gray{+ \frac{1}{5!}K_6 \of{x^5 - 10 x^3 y^2 + 5 x y^4} D_x\delta - \frac{1}{5!}K_6 \of{5 x^4 y - 10 x^2 y^3 + y^5} D_y\delta }\\
        &\quad\quad+ \frac{1}{48}K_6 \of{x^4 - 6 x^2 y^2 + y^4} \of{D_x^2-D_y^2}\delta^2 \gray{- \frac{1}{24} K_6 \of{4 x^3 y - 4 x y^3} D_x D_y \delta^2} \\
        &\quad\quad\gray{+ \frac{1}{36} K_6 \of{x^3 - 3 x y^2} \of{ D_x^3 - 3 D_x D_y^2} \delta^3 - \frac{1}{36} K_6 \of{3 x^2 y - y^3} \of{3 D_x^2 D_y - D_y^3} \delta^3} \\
        &\quad\quad+ \frac{1}{48} K_6 \of{x^2 - y^2} \of{D_x^4 - 6 D_x^2 D_y^2 + D_y^4} \delta^4 \gray{- \frac{1}{6} K_6 x y \of{D_x^3 D_y - D_x D_y^3} \delta^4} \\
        &\quad\quad\gray{+ \frac{1}{120} K_6 x \of{D_x^5 - 10 D_x^3 D_y^2 + 5 D_x D_y^4} \delta^5 - \frac{1}{120} K_6 y \of{5 D_x^4 D_y - 10 D_x^2 D_y^3 + D_y^5} \delta^5}
    \rangle\\
    \quad &\overeq{\cref{eq:CosOddExp}} \quad
        \gray{\langle \frac{1}{6!} K_6 \of{x^6 - 15 x^4 y^2 + 15 x^2 y^4 - y^6} \rangle} \\
        &\quad\quad+ \frac{1}{12}K_6 \of{J_x^2\beta_x^2\langle \cos^4 \phi_x \rangle - 6 J_x\beta_x\langle \cos^2 \phi_x \rangle J_y\beta_y \langle \cos^2 \phi_y \rangle + J_y^2\beta_y^2 \langle \cos^4 \phi_y \rangle } \of{D_x^2-D_y^2}\delta^2 \\
        &\quad\quad+ \frac{1}{48} K_6 \of{2J_x\beta_x\langle \cos^2 \phi_x \rangle - 2J_y\beta_y \langle \cos^2 \phi_y \rangle } \of{D_x^4 - 6 D_x^2 D_y^2 + D_y^4} \delta^4 \\
    \quad &\overeq{\cref{eq:ExpCos}} \quad
        \gray{\langle \frac{1}{6!} K_6 \of{x^6 - 15 x^4 y^2 + 15 x^2 y^4 - y^6} \rangle} \\
        &\quad\quad\gray{+ \frac{1}{32}K_6 \of{J_x^2\beta_x^2 - 4 J_x\beta_x J_y\beta_y  + J_y^2\beta_y^2} \of{D_x^2-D_y^2}\delta^2} \\
        &\quad\quad+ \frac{1}{48} K_6 \of{J_x\beta_x - J_y\beta_y} \of{D_x^4 - 6 D_x^2 D_y^2 +  D_y^4} \delta^4 \\
\end{split}
\end{align}
From this we see, that $Q'$, $Q''$ and $Q'''$ will still depend on either $\delta$, $J_x$ or $J_y$ (or powers of these) and can, with the same arguments as before, be neglected.
The main contribution of a normal dodecapole is hence to $Q''''$: 
\begin{align}
    \frac{\partial^4 \Delta Q_x^{N_6}}{\pardelta^4}
    \quad &\overeq{\cref{eq:tuneChangeChroma}} \quad
    \frac{1}{2\pi} \int_L \frac{\partial^5 \langle N_6\of{x+D_x\delta, y+D_y\delta} \rangle }{\pardelta^4 \; \partial J_x} ds
    \quad \myover{\cref{eq:n6OmomentumExpectation}}{\approx} \quad
    \nosign \frac{1}{4\pi} K_6L \beta_x \of{D_x^4 - 6 D_x^2 D_y^2 + D_y^4}
    \label{eq:fourthchromaXN6}
    \\
    \frac{\partial^4 \Delta Q_y^{N_6}}{\pardelta^4} 
    \quad &\overeq{\cref{eq:tuneChangeChroma}} \quad
    \frac{1}{2\pi} \int_L \frac{\partial^5 \langle S_5\of{x+D_x\delta, y+D_y\delta} \rangle }{\pardelta^4 \; \partial J_y} ds 
    \quad \myover{\cref{eq:n6OmomentumExpectation}}{\approx} \quad
    - \frac{1}{4\pi} K_6^L  \beta_y \of{D_x^4 - 6 D_x^2 D_y^2 + D_y^4}
    \label{eq:fourthchromaYN6}
\end{align}

\subsubsection{Skew Dodecapole}
\label{sec:ChromaFromSkewDodecapole}

\begin{align}
\label{eq:s6OmomentumExpectation}
\begin{split}
    &\langle S_6\of{x+D_x\delta, y+D_y\delta} \rangle \\
    \quad &\myover{\substack{\cref{eq:s6O} \\ + Taylor}}{\approx} \quad
    \langle 
        \gray{- \frac{1}{6!} J_6 \of{6 x^5 y - 20 x^3 y^3 + 6 x y^5}} \\
        &\quad\quad\gray{- \frac{1}{5!}J_6 \of{5 x^4 y - 10 x^2 y^3 + y^5} D_x\delta - \frac{1}{5!}J_6 \of{x^5 - 10 x^3 y^2 + 5 x y^4} D_y\delta}\\
        &\quad\quad\gray{- \frac{1}{48}J_6 \of{4 x^3 y - 4 x y^3} \of{D_x^2-D_y^2}\delta^2} - \frac{1}{24} J_6 \of{x^4 - 6 x^2 y^2 + y^4} D_x D_y \delta^2 \\
        &\quad\quad\gray{- \frac{1}{36} J_6 \of{3 x^2 y - y^3} \of{D_x^3 - 3 D_x D_y^2} \delta^3 -  \frac{1}{36} J_6 \of{x^3 - 3 x y^2} \of{3D_x^2 D_y - D_y^3} \delta^3} \\
        &\quad\quad\gray{- \frac{1}{24} J_6 x y \of{D_x^4 - 6 D_x^2 D_y^2 + D_y^4} \delta^4} - \frac{1}{12} J_6 \of{x^2 - y^2} \of{D_x^3 D_y - D_x D_y^3} \delta^4 \\
        &\quad\quad\gray{- \frac{1}{120} J_6 y \of{D_x^5 - 10 D_x^3 D_y^2 + 5 D_x D_y^4} \delta^5 - \frac{1}{120} J_6 x \of{5 D_x^4 D_y - 10 D_x^2 D_y^3 + D_y^5} \delta^5}
    \rangle\\
    \quad &\overeq{\cref{eq:CosOddExp}} \quad
        - \frac{1}{6}J_6 \of{J_x^2\beta_x^2\langle \cos^4 \phi_x \rangle - 6 J_x\beta_x\langle \cos^2 \phi_x \rangle J_y\beta_y \langle \cos^2 \phi_y \rangle + J_y^2\beta_y^2 \langle \cos^4 \phi_y \rangle } D_x D_y \delta^2 \\
        &\quad\quad- \frac{1}{12} J_6 \of{2J_x\beta_x\langle \cos^2 \phi_x \rangle - 2J_y\beta_y \langle \cos^2 \phi_y \rangle } \of{D_x^3 D_y - D_x D_y^3} \delta^4 \\
    \quad &\overeq{\cref{eq:ExpCos}} \quad
        \gray{+ \frac{1}{16}J_6 \of{J_x^2\beta_x^2 - 4 J_x\beta_x J_y\beta_y  + J_y^2\beta_y^2} \of{D_x^2-D_y^2}\delta^2} \\
        &\quad\quad- \frac{1}{12} J_6 \of{J_x\beta_x - J_y\beta_y} \of{D_x^3 D_y - D_x D_y^3} \delta^4 \\
\end{split}
\end{align}
As in \cref{sec:ChromaFromNormalDodecapole} we see that $Q'$, $Q''$ and $Q'''$ can be neglected and
the main contribution of a skew dodecapole is also to $Q''''$: 
\begin{align}
    \frac{\partial^4 \Delta Q_x^{S_6}}{\pardelta^4}
    \quad &\overeq{\cref{eq:tuneChangeChroma}} \quad
    \frac{1}{2\pi} \int_L \frac{\partial^5 \langle S_6\of{x+D_x\delta, y+D_y\delta} \rangle }{\pardelta^4 \; \partial J_x} ds
    \quad \myover{\cref{eq:s6OmomentumExpectation}}{\approx} \quad
    - \frac{1}{\pi} J_6L \beta_x \of{D_x^3 D_y - D_x D_y^3}
    \label{eq:fourthchromaXS6}
    \\
    \frac{\partial^4 \Delta Q_y^{S_6}}{\pardelta^4} 
    \quad &\overeq{\cref{eq:tuneChangeChroma}} \quad
    \frac{1}{2\pi} \int_L \frac{\partial^5 \langle S_5\of{x+D_x\delta, y+D_y\delta} \rangle }{\pardelta^4 \; \partial J_y} ds 
    \quad \myover{\cref{eq:s6OmomentumExpectation}}{\approx} \quad
    \nosign \frac{1}{\pi} J_6^L  \beta_y \of{D_x^3 D_y - D_x D_y^3}
    \label{eq:fourthchromaYS6}
\end{align}
\section{Outlook}
We hope this note has proven useful to the reader and they could 
verify their own derivations or at least follow along ours and gain some insight 
into the origin behind the different terms of formulas often simply applied.

By no means does this note claim to be a complete collection.
In fact, formulas to calculate chromatic amplitude detuning are already in preparation and will be 
added in future versions.

As accelerator optics become more and more sensitive to 
higher and higher field orders, additional equations may need to be derived to 
account for these fields. 
We hope the reader is now equipped with the knowledge and inspiration to derive the formulas on their own.

\section{Acknowledgements}

The authors would like to thank E. H. Maclean and R. Tom\'as 
for their support and feedback when writing this note.

\FloatBarrier
\include{measurement}

% Bibliography %%%%%%%%%%%%%%%%%%%%%%%%%%%%%%%%%%%%%%%%%%%%%%%%%%%%%%%%%%%%%%%%%%%%%%%%%%%%%%%%%%%% 
\FloatBarrier

\bibliographystyle{siamurl}
\bibliography{library_jdilly.bib}

\end{document}